\documentclass[aps,prl,preprint,superscriptaddress,longbibliography,showkeys]{revtex4-2}
\usepackage[T1]{fontenc}
\usepackage[utf8]{inputenc}
\usepackage{graphicx}
\usepackage{amsmath,amssymb,amsfonts}
\usepackage{multirow}
\usepackage{booktabs}
\usepackage[colorlinks=true,
            linkcolor=black,
            citecolor=cyan,
            urlcolor=black]{hyperref}
\usepackage{xcolor}
\usepackage{textcomp}

\begin{document}
\title{First measurement of kaonic deuterium X-ray transitions}
\author{Massimiliano Bazzi}
\thanks{These authors contributed equally to this work.}
\affiliation{INFN Laboratori Nazionali di Frascati, Frascati, Italy}
\author{Francesco Clozza}
\thanks{These authors contributed equally to this work.}
\affiliation{INFN Laboratori Nazionali di Frascati, Frascati, Italy}
\affiliation{Università degli Studi di Roma Tor Vergata, Dipartimento di Fisica, Roma, Italy}
\author{Carlo Guaraldo}
\thanks{These authors contributed equally to this work.}
\thanks{Deceased.}
\affiliation{INFN Laboratori Nazionali di Frascati, Frascati, Italy}
\author{Mihail Antoniu Iliescu}
\thanks{These authors contributed equally to this work.}
\affiliation{INFN Laboratori Nazionali di Frascati, Frascati, Italy}
\author{Alessandro Scordo}
\thanks{These authors contributed equally to this work.}
\affiliation{INFN Laboratori Nazionali di Frascati, Frascati, Italy}
\author{Francesco Sgaramella}
\thanks{These authors contributed equally to this work.}
\email{Francesco.Sgaramella@lnf.infn.it}
\affiliation{INFN Laboratori Nazionali di Frascati, Frascati, Italy}
\author{Diana Sirghi}
\thanks{These authors contributed equally to this work.}
\affiliation{INFN Laboratori Nazionali di Frascati, Frascati, Italy}
\affiliation{Horia Hulubei National Institute of Physics and Nuclear Engineering (IFIN-HH), Magurele, Romania}
\affiliation{Centro Ricerche Enrico Fermi - Museo Storico della Fisica e Centro Studi e Ricerche "Enrico Fermi", Roma, Italy}
\author{Florin Sirghi}
\thanks{These authors contributed equally to this work.}
\affiliation{INFN Laboratori Nazionali di Frascati, Frascati, Italy}
\affiliation{Horia Hulubei National Institute of Physics and Nuclear Engineering (IFIN-HH), Magurele, Romania}
\author{Johann Zmeskal}
\thanks{These authors contributed equally to this work.}
\thanks{Deceased.}
\affiliation{Stefan Meyer Institute for Subatomic Physics, Vienna, Austria}
\author{Leonardo Abbene}
\affiliation{INFN Laboratori Nazionali di Frascati, Frascati, Italy}
\affiliation{Department of Physics and Chemistry (DiFC)—Emilio Segrè, University of Palermo, Palermo, Italy}
\author{Claude Amsler}
\altaffiliation{Present address: Marietta Blau Institute for Particle Physics, Vienna, Austria}
\affiliation{Stefan Meyer Institute for Subatomic Physics, Vienna, Austria}
\author{Francesco Artibani}
\affiliation{INFN Laboratori Nazionali di Frascati, Frascati, Italy}
\affiliation{Università degli studi di Roma Tre, Dipartimento di Fisica, Roma, Italy}
\author{Giacomo Borghi}
\affiliation{Politecnico di Milano, Dipartimento di Elettronica, Informazione e Bioingegneria, Milano, Italy}
\affiliation{INFN Sezione di Milano, Milano, Italy}
\author{Damir Bosnar}
\affiliation{Department of Physics, Faculty of Science, University of Zagreb, Zagreb, Croatia}
\author{Mario Bragadireanu}
\affiliation{Horia Hulubei National Institute of Physics and Nuclear Engineering (IFIN-HH), Magurele, Romania}
\author{Antonino Buttacavoli}
\affiliation{INFN Laboratori Nazionali di Frascati, Frascati, Italy}
\affiliation{Department of Physics and Chemistry (DiFC)—Emilio Segrè, University of Palermo, Palermo, Italy}
\author{Marco Carminati}
\affiliation{Politecnico di Milano, Dipartimento di Elettronica, Informazione e Bioingegneria, Milano, Italy}
\affiliation{INFN Sezione di Milano, Milano, Italy}
\author{Alberto Clozza}
\affiliation{INFN Laboratori Nazionali di Frascati, Frascati, Italy}
\author{Luca De Paolis}
\affiliation{INFN Laboratori Nazionali di Frascati, Frascati, Italy}
\author{Raffaele Del Grande}
\affiliation{INFN Laboratori Nazionali di Frascati, Frascati, Italy}
\affiliation{Faculty of Nuclear Sciences and Physical Engineering, Czech Technical University in Prague, Prague, Czech Republic}
\author{Kamil Dulski}
\affiliation{INFN Laboratori Nazionali di Frascati, Frascati, Italy}
\affiliation{Faculty of Physics, Astronomy, and Applied Computer Science, Jagiellonian University, Kraków, Poland}
\affiliation{Center for Theranostics, Jagiellonian University, Kraków, Poland}
\author{Carlo Fiorini}
\affiliation{Politecnico di Milano, Dipartimento di Elettronica, Informazione e Bioingegneria, Milano, Italy}
\affiliation{INFN Sezione di Milano, Milano, Italy}
\author{Ivica Friščić}
\affiliation{Department of Physics, Faculty of Science, University of Zagreb, Zagreb, Croatia}
\author{Masa Iwasaki}
\affiliation{RIKEN, Tokyo, Japan}
\author{Aleksander Khreptak}
\affiliation{INFN Laboratori Nazionali di Frascati, Frascati, Italy}
\affiliation{Faculty of Physics, Astronomy, and Applied Computer Science, Jagiellonian University, Kraków, Poland}
\affiliation{Center for Theranostics, Jagiellonian University, Kraków, Poland}
\author{Simone Manti}
\affiliation{INFN Laboratori Nazionali di Frascati, Frascati, Italy}
\author{Johann Marton}
\altaffiliation{Present address: Technische Universität Wien, Atominstitut, Vienna, Austria}
\affiliation{Stefan Meyer Institute for Subatomic Physics, Vienna, Austria}
\author{Catia Milardi}
\affiliation{INFN Laboratori Nazionali di Frascati, Frascati, Italy}
\author{Pawel Moskal}
\affiliation{Faculty of Physics, Astronomy, and Applied Computer Science, Jagiellonian University, Kraków, Poland}
\affiliation{Center for Theranostics, Jagiellonian University, Kraków, Poland}
\author{Fabrizio Napolitano}
\altaffiliation{Present address: Dipartimento di Fisica e Geologia, Università degli studi di Perugia, Perugia, Italy; INFN Sezione di Perugia, Perugia, Italy}
\affiliation{INFN Laboratori Nazionali di Frascati, Frascati, Italy}
\author{Hiroaki Ohnishi}
\affiliation{Research Center for Accelerator and Radioisotope Science (RARiS), Tohoku University, Sendai, Japan}
\author{Kristian Piscicchia}
\affiliation{INFN Laboratori Nazionali di Frascati, Frascati, Italy}
\affiliation{Centro Ricerche Enrico Fermi - Museo Storico della Fisica e Centro Studi e Ricerche "Enrico Fermi", Roma, Italy}
\author{Fabio Principato}
\affiliation{INFN Laboratori Nazionali di Frascati, Frascati, Italy}
\affiliation{Department of Physics and Chemistry (DiFC)—Emilio Segrè, University of Palermo, Palermo, Italy}
\author{Michał Silarski}
\affiliation{Faculty of Physics, Astronomy, and Applied Computer Science, Jagiellonian University, Kraków, Poland}
\author{Magdalena Skurzok}
\affiliation{INFN Laboratori Nazionali di Frascati, Frascati, Italy}
\affiliation{Faculty of Physics, Astronomy, and Applied Computer Science, Jagiellonian University, Kraków, Poland}
\affiliation{Center for Theranostics, Jagiellonian University, Kraków, Poland}
\author{Antonio Spallone}
\affiliation{INFN Laboratori Nazionali di Frascati, Frascati, Italy}
\author{Kairo Toho}
\affiliation{INFN Laboratori Nazionali di Frascati, Frascati, Italy}
\affiliation{Research Center for Accelerator and Radioisotope Science (RARiS), Tohoku University, Sendai, Japan}
\author{Marlene T\"uchler}
\affiliation{Stefan Meyer Institute for Subatomic Physics, Vienna, Austria}
\author{Oton Vazquez Doce}
\affiliation{INFN Laboratori Nazionali di Frascati, Frascati, Italy}
\author{Eberhard Widmann}
\altaffiliation{Present address: Marietta Blau Institute for Particle Physics, Vienna, Austria}
\affiliation{Stefan Meyer Institute for Subatomic Physics, Vienna, Austria}
\author{Catalina Curceanu}
\affiliation{INFN Laboratori Nazionali di Frascati, Frascati, Italy}
\begin{abstract}
The study of the strong interaction among hadrons at low energies remains one of the key challenges in fundamental physics because of its non-perturbative nature, which makes theoretical descriptions strongly dependent on experimental input. Although substantial progress has been made for systems involving up and down quarks, theoretical models in the strangeness sector continue to face limitations due to the lack of experimental data. Kaonic atoms provide a powerful tool to study the low-energy strong interaction with strangeness through the energy shifts and widths induced on their lowest atomic levels. In this context, kaonic deuterium X-ray spectroscopy has long represented one of the major open challenges in hadronic-atom physics because of its extremely low X-ray yield. This measurement is particularly important because it gives access to the poorly constrained $K^-n$ interaction at threshold. Here, we report the first observation of kaonic deuterium X-ray transitions, performed with the SIDDHARTA-2 experiment at the DA$\Phi$NE collider. We determine the strong-interaction shift and width of the $1s$ level to be $\varepsilon_{1s}=-810.9\pm24.5\,(\mathrm{stat})\pm2.1\,(\mathrm{syst})\,\mathrm{eV}$ and $\Gamma_{1s}=812\pm97\,(\mathrm{stat})\pm33\,(\mathrm{syst})\,\mathrm{eV}$, respectively. This measurement constitutes the most precise experimental determination of the $K^-d$ strong interaction at threshold and allows discrimination among competing theoretical models. Combined with the kaonic hydrogen measurement, this result provides the experimental input required to determine the isospin-dependent $K^-N$ scattering lengths, with implications for the description of the nature of the molecular interpretation of the $\Lambda(1405)$, and neutron-rich matter.
\end{abstract}
\maketitle

\section{Introduction}
The strong interaction, described within the Standard Model by Quantum Chromodynamics (QCD), is responsible for binding quarks and gluons into hadrons and, ultimately, for the structure and stability of matter in the universe. While QCD is well tested at high energies, where perturbative methods are successfully applied, its low-energy domain remains a central open problem in nuclear and particle physics.
In this regime, the strong coupling increases, leading to non-perturbative phenomena that require precise experimental data to constrain the theoretical descriptions.
In this context, hadronic atoms provide a powerful laboratory to access the strong interaction at very low energy \cite{Gasser:2007zt}. In these systems, a negatively charged hadron replaces an atomic electron, and, following a series of de-excitation processes, cascades towards lower-lying states, emitting characteristic X-rays.

For atomic states that approach the nuclear region to within a few tens of femtometres, the strong interaction between the hadron and the nucleus induces an energy shift ($\varepsilon$) and a broadening ($\Gamma$) of the energy levels with respect to the purely electromagnetic values.
The corresponding hadron–nucleus relative energy is only a few keV, placing the system close to threshold.
Precision X-ray spectroscopy of these transitions therefore provides direct access to the non-perturbative regime of the strong hadron–nucleon interaction.\\
Among hadronic atoms, kaonic atoms are of particular interest because they probe the strangeness sector \cite{Curceanu:2026zjg}.
The underlying antikaon–nucleon ($K^-N$) dynamics is shaped by the $\Lambda(1405)$ resonance, which lies just below the $K^-N$ threshold and couples strongly to the $K^-N$ and $\Sigma\pi$ channels \cite{Dalitz:1959dn,Hyodo:2011ur}.
Rather than being described as a conventional three-quark baryon, the $\Lambda(1405)$ is widely interpreted as a dynamically generated state with a dominant meson--baryon molecular component \cite{Mai:2020ltx}. Establishing its structure and accurately describing its properties are therefore of broad relevance to hadron spectroscopy, particularly in view of the growing number of near-threshold states that are interpreted or predicted as hadronic molecules \cite{Guo:2017jvc}.\\
As a consequence, a reliable description of the low-energy $K^-N$ interaction requires non-perturbative coupled-channel approaches based on chiral SU(3) effective field theory \cite{Ikeda:2012au,Guo:2012vv,Mai:2014xna,Cieply:2011nq,Feijoo:2018den}. These descriptions remain strongly dependent on experimental constraints, especially in the near-threshold region.

The SIDDHARTA measurement of the strong interaction induced shift and width of kaonic hydrogen $1s$ level determined the $K^-p$ scattering length, providing a key constraint on theoretical descriptions of the low-energy $K^-N$ interaction \cite{SIDDHARTA:2011dsy}. 
However, a complete determination of the $K^-N$ interaction requires knowledge of both isospin components. While the $K^-p$ system involves a combination of the isospin $I=0$ and $I=1$ amplitudes, the $K^-n$ interaction is purely $I=1$. Nowadays, theoretical models \cite{Ikeda:2012au,Guo:2012vv,Mai:2014xna,Cieply:2011nq,Feijoo:2018den} provide significantly divergent predictions for the $K^-n$ scattering length, due to the lack of experimental data. \\
X-ray spectroscopy of kaonic deuterium ($K^-d$) provides access to the strong interaction shift and width of the $1s$ level and hence to the $K^-d$ scattering length (see End Matter). Together with the kaonic hydrogen result, this measurement makes it possible to determine the isoscalar and isovector $K^-N$ scattering lengths, thereby providing the missing experimental constraint on the poorly known $K^-n$ interaction.\\
On the theoretical side, several approaches have been developed to describe the $K^-d$ system, differing in the treatment of the three-body dynamics and the kaon--nucleon interaction. These range from simplified schemes, such as the fixed-centre approximation (FCA) \cite{Kamalov:2000iy,Ramos:2025ibe}, to more complex three-body calculations \cite{Revai:2016muw,Shevchenko:2011ce,Mizutani:2012gy}, and lead to substantially different predictions for the strong interaction shift and width of the $1s$ atomic level. A measurement of kaonic deuterium therefore provides the experimental input required to anchor theoretical models describing the $K^-d$ interaction.

Knowledge of both isospin components of the $K^-N$ interaction is also important for describing the in-medium kaon interaction in neutron-rich matter \cite{Friedman:2012pc}. In models of neutron-star matter, an attractive in-medium kaon interaction may lead to kaon condensation \cite{DePietri:2019khb,Tolos:2020aln,Merafina:2020ffb}. Such a condensate changes the particle composition and can soften the equation of state, thereby affecting neutron-star masses and radii \cite{Mishra:2010Kaons,Thapa:2020Kaons,Tolos:2020aln}.\\
Kaonic atom spectroscopy constrains the interaction at threshold but does not directly determine the energy dependence of the sub-threshold amplitudes. Nevertheless, theoretical descriptions of these amplitudes must reproduce the measured threshold observables. These constraints are therefore essential for predictions of the proposed $K^-pp$ quasi-bound state and other $K^-NN$ systems \cite{Yamazaki:2007yc,Akaishi:2003jk}, whose binding energies and widths depend on the sub-threshold $K^-N$ amplitudes \cite{Dote:2008Kpp,Dote:2018Kpp,Shevchenko:2016wnu}.

Kaonic deuterium remained the last unexplored light kaonic atom after about five decades of experimental attempts. Its expected X-ray yield to the $1s$ level is about an order of magnitude below that of kaonic hydrogen \cite{SIDDHARTA:2013ftj}, while its larger predicted width ($0.6\div2$ keV) further reduces the signal-to-background ratio. Reaching a precision comparable to the SIDDHARTA kaonic hydrogen measurement \cite{SIDDHARTA:2011dsy}, sufficient to discriminate among theoretical models, therefore required the development of a new experimental apparatus capable of achieving a substantially improved signal-to-background ratio.
This has driven the design of the SIDDHARTA-2 apparatus \cite{Sirghi:2023wok}, which combines a new generation of large-area spectroscopic X-ray Silicon Drift Detectors (SDDs) with enhanced energy and time resolution, an upgraded kaon trigger system, and dedicated veto systems for background suppression. \\

In this work we report the first X-ray spectroscopy of kaonic deuterium performed by the SIDDHARTA-2 experiment at the DA$\Phi$NE collider. By measuring the X-ray transitions to the ground state, we determined the strong interaction induced shift ($\varepsilon_{1s}$) and width ($\Gamma_{1s}$) on the kaonic deuterium $1s$ level, providing the most precise experimental observation of the $K^-d$ interaction at threshold energy.

\section{Kaonic deuterium X-ray spectroscopy}\label{sec2}
A kaonic deuterium atom is formed when a negatively charged kaon enters a gaseous target and loses its kinetic energy through ionization and excitation of the surrounding atoms. The kaon is then captured electromagnetically by a deuterium atom, replacing its electron and initially occupying a highly excited atomic state ($n \simeq 28$). This high initial quantum number results from the kaon mass being approximately 1,000 times the electron mass. The kaonic atom subsequently de-excites towards the ground state through collisional processes and radiative transitions. The cascade involves X-ray and Auger emission as well as non-radiative mechanisms, such as Coulomb de-excitation \cite{Jensen:2002wq,Jensen:2002wr}.
By measuring both the energy and line shape of X-ray transitions to the $1s$ level, the strong interaction induced shift ($\varepsilon_{1s}$) and width ($\Gamma_{1s}$) can be determined, providing direct access to the $K^-d$ strong interaction at threshold energy.

\begin{figure}[htbp]
    \centering
    \includegraphics[width=0.8\columnwidth]{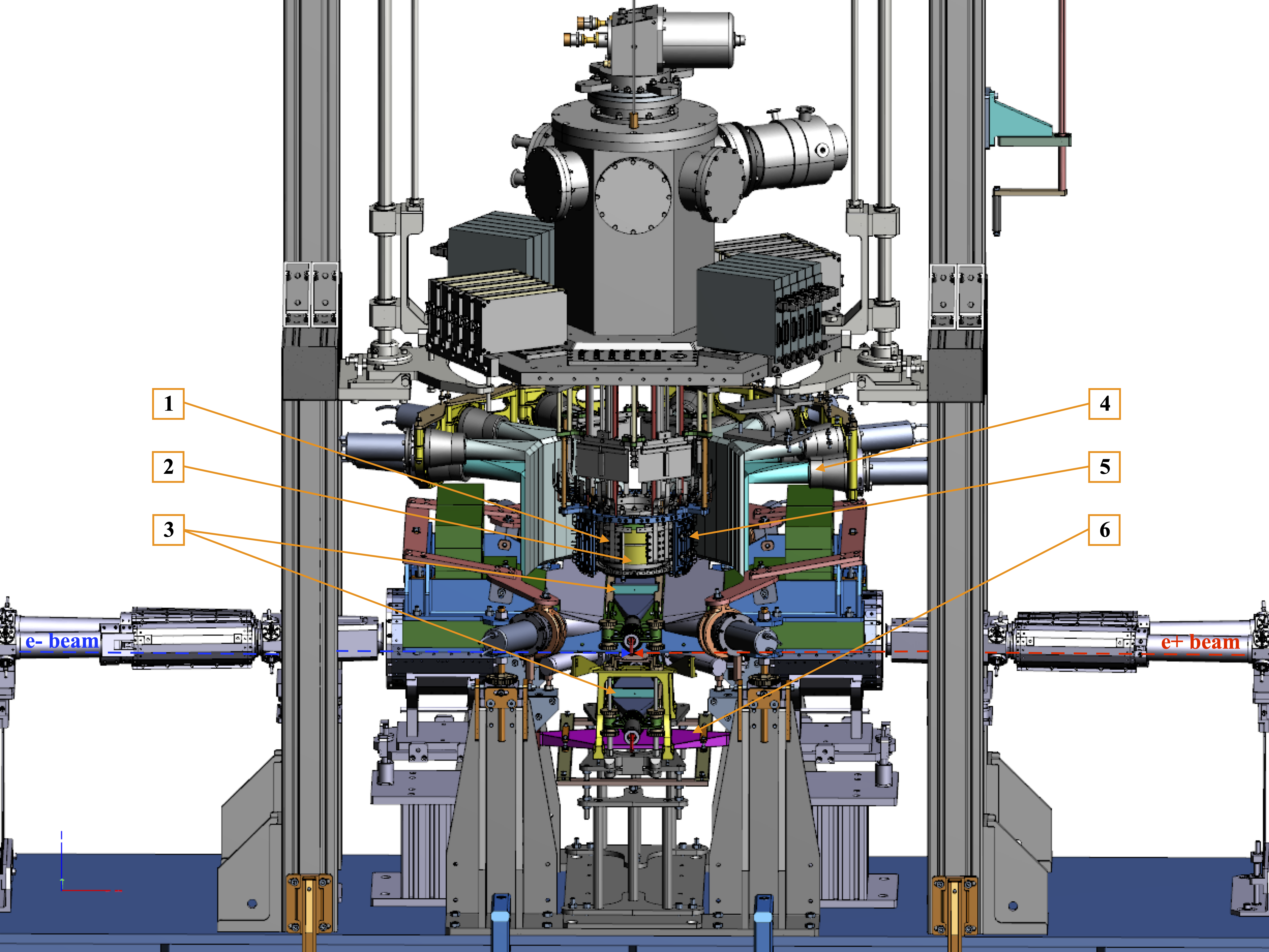}
    \caption{\textbf{Schematic layout of the SIDDHARTA-2 apparatus at the DA$\Phi$NE collider.} Front cutaway view of the apparatus positioned above the DA$\Phi$NE interaction point; the $e^-$ and $e^+$ beam lines are shown in blue and red, respectively. The numbered components are: (1) the array of 384 Silicon Drift Detectors used for X-ray spectroscopy; (2) the cryogenic gaseous-deuterium target cell; (3) the kaon trigger, comprising top and bottom plastic-scintillator counters; (4) the Veto-1 system; (5) the Veto-2 system; and (6) the Veto-3 system, which acts as the charged-kaon veto.}
    \label{fig:4}
\end{figure}

The kaonic deuterium measurement was performed with the SIDDHARTA-2 experiment at the DA$\Phi$NE $e^+e^-$ collider \cite{Zobov:2018yxf,Zobov:2010zza} of INFN-LNF during the 2023–2024 data-taking campaign. The DA$\Phi$NE collider delivers low-momentum ($\sim 127~\mathrm{MeV}/c$) $K^+K^-$ pairs from $\phi$-meson decays, offering a unique environment for the formation and study of kaonic atoms. The collider configuration was specifically optimized to enhance the kaon yield while suppressing the background \cite{Milardi:2018sih,Milardi:2021khj}, achieving an improvement of about a factor of three in the signal-to-background ratio compared to the SIDDHARTA kaonic hydrogen measurement \cite{Milardi:2024efr}. The SIDDHARTA-2 apparatus (Fig. \ref{fig:4}) is installed above the DA$\Phi$NE interaction point, where back-to-back $K^+K^-$ pairs are tagged by two scintillator counters forming the kaon trigger. 
Kaons emitted towards the apparatus enter the vacuum chamber through a $125~\mu\mathrm{m}$-thick Kapton window and reach the cryogenic gaseous-deuterium target, surrounded by an array of Silicon Drift Detectors with a total active area of $245~\mathrm{cm}^2$. The inner Veto-2 and outer Veto-1 systems, both composed of plastic scintillators, are installed inside and outside the vacuum chamber, respectively, and together form a two-layer veto barrel around the target. Veto-3, the charged-kaon veto, is positioned below the lower trigger counter. A description of the experimental apparatus is given in Ref.
\cite{Sirghi:2023wok}.

The target density was chosen to balance efficient kaon stopping against the suppression of the X-ray yield by Stark mixing at high densities \cite{Jensen:2002wr}. The deuterium target was operated at $25~\mathrm{K}$, with a density of $2.28~\mathrm{g/l}$ (1.4\% of the liquid deuterium density). These conditions maximize the X-ray output by retaining a sizeable radiative yield while stopping a sufficient number of kaons.
The X-rays emitted from the kaonic deuterium atoms were detected using 384 SDDs \cite{Miliucci:2021wbj,Miliucci:2022lvn}, specifically developed for kaonic atoms X-ray spectroscopy and arranged around the deuterium target. The data were collected with a total integrated luminosity of 1.1 fb$^{-1}$ ($\sim 11 \times 10^7$ triggered K$^+$K$^-$ pairs).

The electromagnetic background, arising from beam losses (Touschek effect) and the resulting electromagnetic showers, is largely asynchronous with kaon production and is suppressed by a factor of $10^4$ by using the timing information provided by the trigger system and the SDDs. In contrast, hadronic background originates from kaon interactions in the apparatus and cannot be removed by the SDD–trigger coincidence alone. Its suppression was one of the key challenges of the kaonic deuterium measurement which motivated the development of three veto systems, a major innovation of the SIDDHARTA-2 apparatus with respect to SIDDHARTA. A detailed description of the event selection procedure is provided in End Matter.

\begin{figure}[htbp]
    \centering
    \includegraphics[width=0.8\columnwidth]{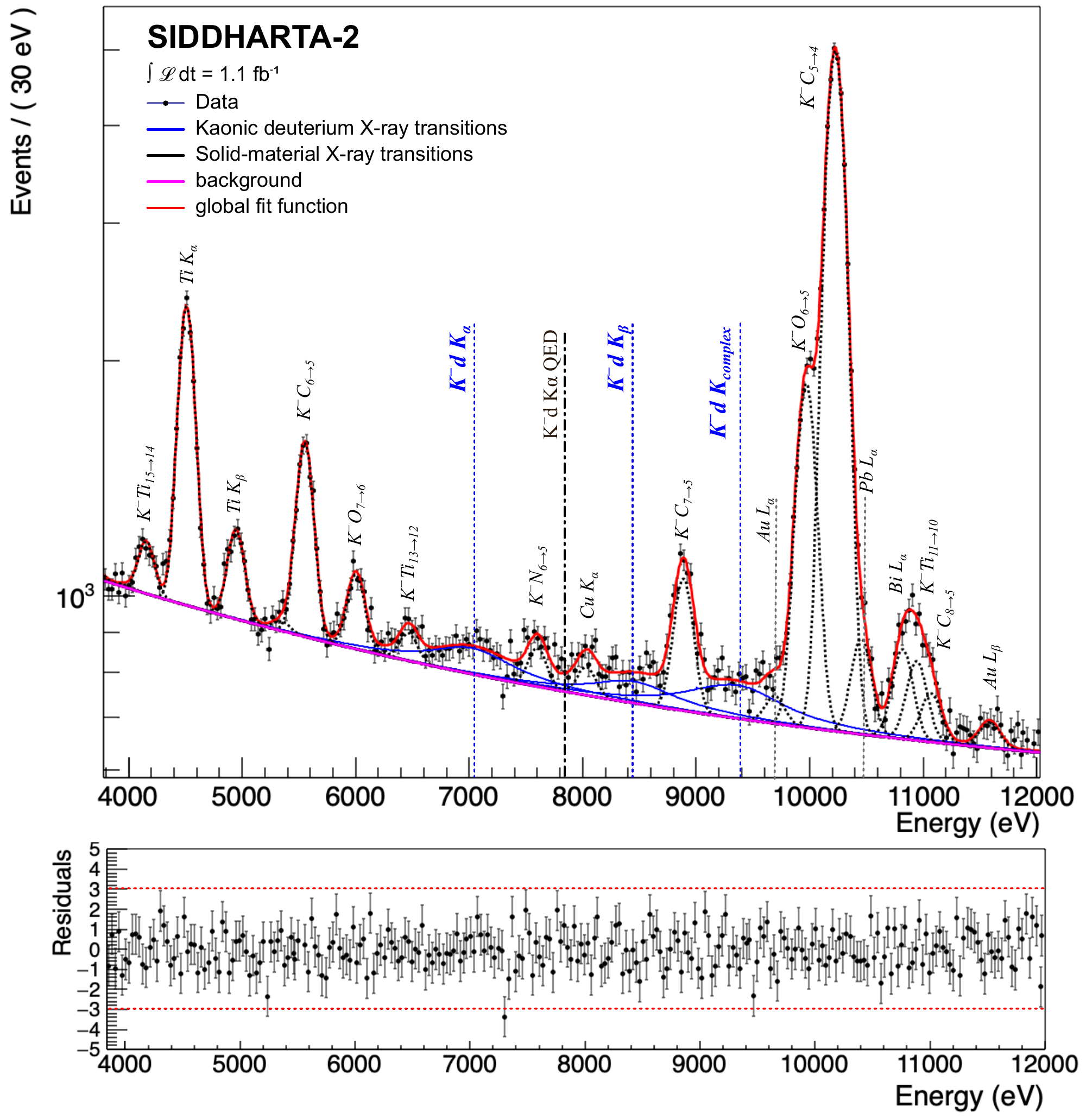}
    \caption{\textbf{Kaonic deuterium X-ray spectrum and fit.} The main panel shows the measured X-ray data points with statistical error bars and the overall fit function, including the individual contributions from the kaonic deuterium transitions, kaonic atoms formed in the surrounding solid materials, primarily Kapton, and fluorescence lines from the Pb shielding, Au, Bi and Ti. The vertical dot--dashed line indicates the purely electromagnetic (QED) prediction for the $K_{\alpha}$ (2p$\to$1s) transition. The lower panel displays the pull distribution (residuals divided by statistical uncertainties) between the data and the fit, providing a visual assessment of the fit quality. The fit gives $\chi^2/\mathrm{ndf}=1.12$.}
    \label{fig:2}
\end{figure}

The measured kaonic deuterium energy spectrum is shown in Fig. \ref{fig:2}. An extended maximum-likelihood fit was performed to extract the transition energies, simultaneously modelling the signal and background components. The kaonic deuterium transitions were modelled with Voigt functions, where the Gaussian term accounts for the detector resolution and the Lorentzian term describes the intrinsic broadening induced by the strong interaction on the $1s$ level. The transition energies were parametrized as $E = E_{\mathrm{QED}} + \varepsilon_{1s}$, with $E_{\mathrm{QED}}$ fixed to the electromagnetic values calculated from the Klein--Gordon equation, including vacuum-polarization and recoil corrections \cite{Santos:2004bw,Karshenboim:2006zz,Karshenboim:2005am} (Table \ref{tab:1}). 
As the strong interaction, in kaonic deuterium, significantly affects only the ground level, the shift $\varepsilon_{1s}$ and width $\Gamma_{1s}$ parameters were common to all transitions to the $1s$ level.
The fit model included the $K_{\alpha}$ ($2p \rightarrow 1s$), $K_{\beta}$ ($3d \rightarrow 1s$) and higher transitions from $4f \rightarrow 1s$ to $6h \rightarrow 1s$, collectively referred to as $K_{\mathrm{complex}}$.
The possible contribution of transitions from higher energy levels was accounted for in the evaluation of the systematic uncertainties.\\
Additional lines from kaonic atoms formed in the target cell's Kapton entrance window ($\mathrm{C}_{22}\mathrm{H}_{10}\mathrm{N}_{2}\mathrm{O}_{5}$) and in the materials of the support structure, as well as fluorescence lines from setup elements, were described with Gaussian functions.
The fit was performed over the 4-12 keV energy range.

The resulting strong interaction induced shift and width of the kaonic deuterium $1s$ level are:
\begin{equation*}
\varepsilon_{1s} = -810.9 \pm 24.5~(\mathrm{stat}) \pm 2.1~(\mathrm{syst})~\mathrm{eV} 
\end{equation*}
\begin{equation*}
\Gamma_{1s} = 812 \pm 97~(\mathrm{stat}) \pm 33~(\mathrm{syst})~\mathrm{eV}
\end{equation*}
The measured negative shift and large width indicate a repulsive-type and strongly absorptive $K^-d$ interaction, providing quantitative experimental access to the low-energy strong interaction in the strangeness sector.

\begin{table}[htbp]
    \centering
    \caption{Electromagnetic (QED) transition energies for kaonic deuterium calculated from the Klein--Gordon equation, including vacuum-polarization and recoil corrections \cite{Santos:2004bw,Karshenboim:2006zz,Karshenboim:2005am}.}
    \label{tab:1}
    \begin{tabular}{lc}
    \hline
    Transition & $E_{\mathrm{QED}}$ (eV) \\
    \hline
    $K_{\alpha}$ ($2p \rightarrow 1s$) & 7834.0 \\
    $K_{\beta}$ ($3d \rightarrow 1s$) & 9280.2 \\
    $4f \rightarrow 1s$ & 9786.2 \\
    $5g \rightarrow 1s$ & 10020.4 \\
    $6h \rightarrow 1s$ & 10147.6 \\
    \hline
    \end{tabular}
\end{table}

\section{Discussion}\label{sec3}
\begin{figure}[t]
    \centering
    \includegraphics[width=0.8\columnwidth]{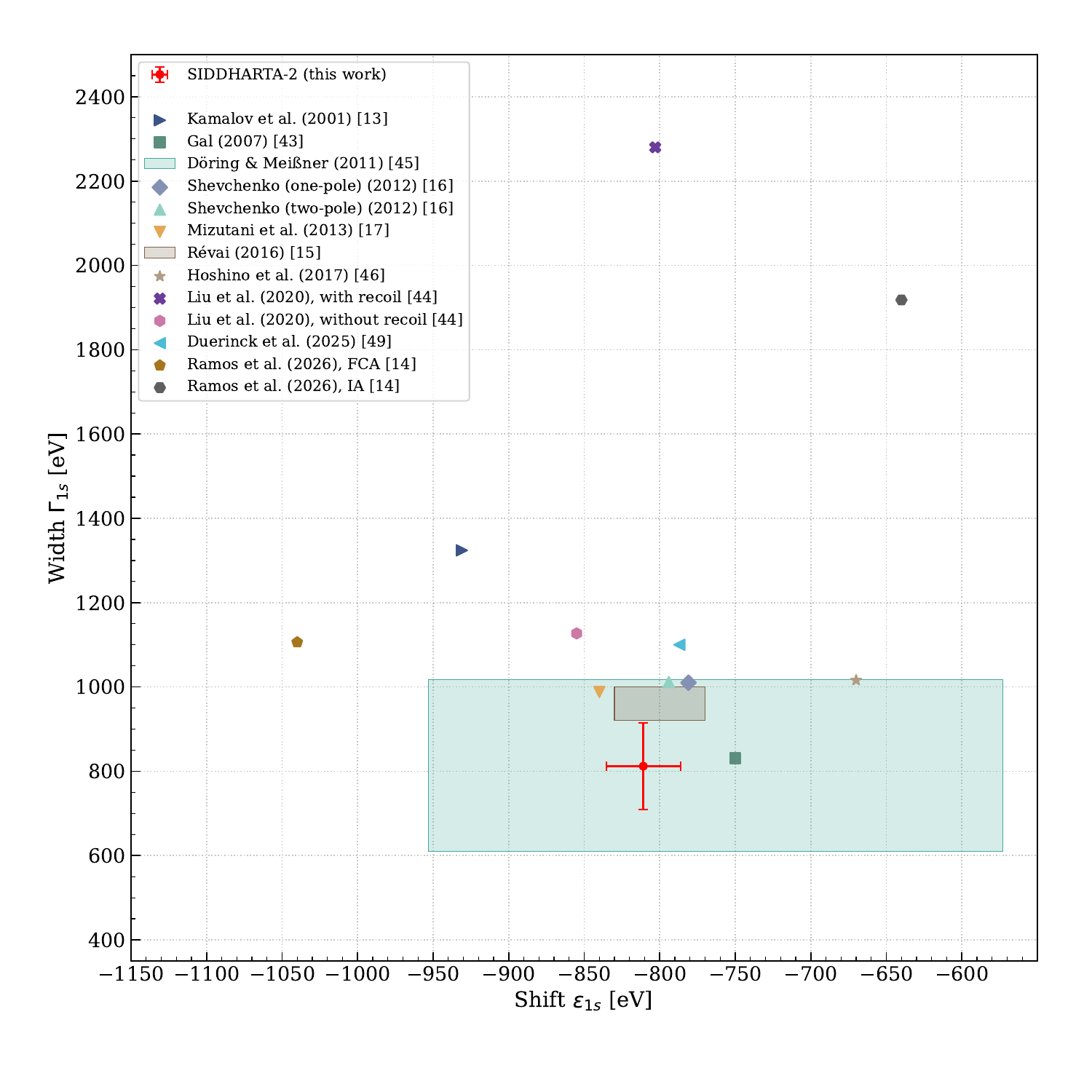}
    \caption{\textbf{Measured SIDDHARTA-2 values (red dot with error bars) of kaonic deuterium $1s$ level shift and width compared with theoretical predictions.} The error bars represent the statistical and systematic uncertainties combined in quadrature. Theoretical models \cite{Revai:2016muw,Shevchenko:2011ce,Mizutani:2012gy,Ramos:2025ibe,Kamalov:2000iy,Gal:2006cw,Liu:2020foc,Hoshino:2017mty,Duerinck:2025pbj,Doring:2011xc} are indicated by different symbols; when theoretical uncertainties are provided, they are shown as shaded rectangles. For the models \cite{Kamalov:2000iy,Gal:2006cw,Mizutani:2012gy,Doring:2011xc}, the $K^-d$ scattering lengths were converted to shifts and widths using the summed-up Deser-type formula \cite{Shevchenko:2021swf}.}
    \label{fig:3}
\end{figure}

Fig. \ref{fig:3} compares the measured $K^-d$ $1s$ level shift and width with several theoretical models. 
Among these, models that treat kaonic deuterium as a dynamical three-body system within full Faddeev frameworks provide the most consistent description of both observables. These include calculations based on Coulomb Sturmian expansions \cite{Revai:2016muw}, phenomenological coupled-channel $K^-NN$--$\pi\Sigma N$ interactions \cite{Shevchenko:2011ce}, and chirally motivated $K^-N$ amplitudes constrained by kaonic hydrogen data \cite{Mizutani:2012gy}. 
The specific impulse approximation (IA) and fixed-centre approximation (FCA) calculations \cite{Kamalov:2000iy,Ramos:2025ibe} shown in Fig. \ref{fig:3} do not reproduce the measured values of both observables simultaneously. Their differences from the data depend on both the elementary $K^-N$ input and the treatment of three-body dynamics, in which aspects of nucleon motion, multiple scattering, or coupled-channel effects are simplified or neglected, depending on the approximation.
Calculations incorporating additional multiple-scattering effects move closer to the experimental result \cite{Gal:2006cw}. A particularly interesting case is provided by the Hamiltonian effective field theory approach of Ref.~\cite{Liu:2020foc}, in which the predicted recoil-enhanced width is almost three times larger than the measured value, whereas neglecting recoil yields a result closer to the data.

The measured shift and width thus provide an experimental benchmark for refining theoretical descriptions of the $K^-d$ strong interaction at threshold, which are needed to extract the isospin-dependent $K^-N$ scattering lengths. The SIDDHARTA kaonic hydrogen measurement \cite{SIDDHARTA:2011dsy} constrains one combination of the isoscalar and isovector scattering lengths, $a_0$ and $a_1$. Kaonic deuterium provides a second independent constraint on a different combination of the two (see End Matter). Their extraction is not a simple linear inversion, because the $K^-d$ scattering length also contains contributions from charge exchange, multiple scattering, nucleon recoil and other three-body effects. When combining the hydrogen and deuterium measurements, all these contributions must be consistently included within an appropriate theoretical treatment \cite{Doring:2011xc,Hoshino:2017mty}.\\
To provide the $K^-d$ input required for this combined determination, we convert the measured $1s$ level shift and width into the complex scattering length using the summed-up Deser formula \cite{Shevchenko:2021swf}, obtaining:
$
a_{K^-d}=(-1.57\pm0.07~(\mathrm{stat})\pm0.01~(\mathrm{syst}))+i(1.11\pm0.13~(\mathrm{stat})\pm0.04~(\mathrm{syst}))~\mathrm{fm} 
$.
An independent determination of the $K^-d$ scattering length was recently reported by the ALICE Collaboration using femtoscopy \cite{ALICE:2026pxr}:
$
\left(-1.44\pm 0.15~(\mathrm{stat})\pm 0.10~(\mathrm{syst})\right)
+i\left(1.34\pm0.33~(\mathrm{stat}){}^{+0.21}_{-0.15}~(\mathrm{syst})\right)~\mathrm{fm}.
$
Our result is consistent with the ALICE determination, while reducing the uncertainty by a factor of approximately 2.5, providing the most precise determination of the $K^-d$ strong interaction at threshold. \\
Previous studies showed that even a measurement of the kaonic deuterium shift with an accuracy of about $25\%$ would provide a stronger constraint on the isovector component of $K^-N$ than the one obtained using the kaonic hydrogen alone \cite{Hoshino:2017mty}. The present experimental uncertainty, approximately $3\%$ for the shift, substantially improves upon this benchmark.

\section{Outlook}\label{sec4} 
The present measurement also provides new constraints for studies of the $\Lambda(1405)$. By determining the $K^-p$ scattering length, the SIDDHARTA kaonic hydrogen measurement reduced the range of pole positions compatible with coupled-channel analyses \cite{SIDDHARTA:2011dsy,Ikeda:2012au,Mai:2020ltx}. However, hydrogen alone constrains a combination of $a_0$ and $a_1$, leaving an ambiguity reflected in the spread of predictions for the $K^-n$ amplitude \cite{Cieply:2016jby}. The complementary deuterium constraint will help disentangle these components and determine the isoscalar amplitude more precisely. Since the $\Lambda(1405)$ poles are obtained by analytically continuing this amplitude, the improved threshold constraint is expected to narrow the allowed pole positions, particularly for the higher-mass, narrower pole, which lies close to the $K^-N$ threshold and couples predominantly to this channel.

The same isospin-dependent scattering lengths are also key inputs for predictions of $K^-pp$ quasi-bound states and dense matter. Within the field-picture ansatz of Ref.~\cite{Dote:2018Kpp}, replacing the older scattering-length constraints with SIDDHARTA kaonic hydrogen constrained input changed the predicted binding-energy interval from 19–37 MeV to 14–28 MeV, reducing its span by approximately 22$\%$. The present kaonic deuterium result extends this experimental constraint to the isovector sector, enabling an analogous reassessment of the sub-threshold amplitudes entering $K^-pp$ calculations. 

Beyond few-body systems, the kaonic deuterium measurement is also relevant to in-medium kaon interaction in neutron-rich matter, where the leading-density $K^-$ self-energy depends on the $K^-p$ and $K^-n$ amplitudes. 
Dense-matter calculations connect low-energy kaon--nucleon information to the in-medium $K^-$ interaction through different implementations, including chiral contact terms calibrated to scattering lengths and optical potentials constrained by kaonic atom data \cite{Mishra:2010Kaons,Thapa:2020Kaons,Wang:2026Kaons}. These calculations generally place the onset of kaon condensation at densities of a few times nuclear saturation density ($n_0$). In the interaction and matter-composition scenarios explored in Ref. \cite{Wang:2026Kaons}, it spans approximately $2$--$8\,n_0$.
This range determines whether kaon condensation can occur over a broad range of neutron-star masses or only in configurations reaching the highest central densities, with implications for stellar composition and the equation of state. The present work will therefore allow the $K^-n$ interaction used in these calculations to be constrained experimentally, enabling a reassessment of the in-medium $K^-$ energy and the onset of kaon condensation.
\section{Summary}\label{sec5}
In this work we reported, for the first time, the measurement of the strong interaction induced shift and width on the kaonic deuterium $1s$ level, closing a long-standing experimental gap in low-energy strangeness physics and providing a benchmark that discriminates among competing $K^-d$ models. Together with kaonic hydrogen data, these observables supply the complementary constraints needed to determine the isoscalar and isovector $K^-N$ scattering lengths, with implications for the $\Lambda(1405)$, $K^-pp$ quasi-bound states and the in-medium kaon interaction in neutron-rich matter.\\
Kaonic deuterium establishes the most precise experimental reference for the interaction of an antikaon with a two-nucleon system at threshold. 
Extending these measurements to heavier kaonic atoms will probe how the interaction evolves with increasing nuclear complexity and constrain kaon--multinucleon dynamics.
The experimental performance demonstrated here provides a benchmark for designing this programme of precision spectroscopy, opening a route to systematic studies of the low-energy strong interaction with strangeness across increasingly complex nuclear systems.\\

\section*{Acknowledgments}
\noindent
We dedicate this work to the memory of our colleagues Carlo Guaraldo and Johann Zmeskal, whose contributions have been essential to the development of the SIDDHARTA-2 experiment and the success of the kaonic deuterium measurement. This work would not have been possible without them.\\
We thank C. Capoccia from LNF-INFN and H. Schneider, L. Stohwasser, and D. Pristauz-Telsnigg from Stefan Meyer-Institut for their fundamental contribution in designing and building the SIDDHARTA-2 setup. We thank as well the INFN, INFN-LNF and the DA$\Phi$NE staff for the excellent working conditions and permanent support. 
Part of this work was supported by the INFN (KAONNIS Project); Austrian Science Fund (FWF): [P24756-N20 and P33037-N]; the Croatian Science Foundation under the project IP-2022-10-3878; the EU STRONG-2020 project (Grant Agreement No. 824093); the EU Horizon 2020 project under the MSCA (Grant Agreement 754496); the Japan Society for the Promotion of Science JSPS KAKENHI Grant No. JP18H05402; the SciMat and qLife Priority Research Areas budget under the program Excellence Initiative - Research University at the Jagiellonian University, and the Polish National Agency for Academic Exchange (Grant No. PPN/BIT/2021/1/00037); the EU Horizon 2020 research and innovation programme under project OPSVIO (Grant Agreement No. 101038099); the Italian Ministry for University and Research (MUR), under PRIN 2022 PNRR project CUP: B53D23024100001.

\bibliography{sn-bibliography}

\begin{thebibliography}{57}%
\makeatletter
\providecommand \@ifxundefined [1]{%
 \@ifx{#1\undefined}
}%
\providecommand \@ifnum [1]{%
 \ifnum #1\expandafter \@firstoftwo
 \else \expandafter \@secondoftwo
 \fi
}%
\providecommand \@ifx [1]{%
 \ifx #1\expandafter \@firstoftwo
 \else \expandafter \@secondoftwo
 \fi
}%
\providecommand \natexlab [1]{#1}%
\providecommand \enquote  [1]{``#1''}%
\providecommand \bibnamefont  [1]{#1}%
\providecommand \bibfnamefont [1]{#1}%
\providecommand \citenamefont [1]{#1}%
\providecommand \href@noop [0]{\@secondoftwo}%
\providecommand \href [0]{\begingroup \@sanitize@url \@href}%
\providecommand \@href[1]{\@@startlink{#1}\@@href}%
\providecommand \@@href[1]{\endgroup#1\@@endlink}%
\providecommand \@sanitize@url [0]{\catcode `\\12\catcode `\$12\catcode
  `\&12\catcode `\#12\catcode `\^12\catcode `\_12\catcode `\%12\relax}%
\providecommand \@@startlink[1]{}%
\providecommand \@@endlink[0]{}%
\providecommand \url  [0]{\begingroup\@sanitize@url \@url }%
\providecommand \@url [1]{\endgroup\@href {#1}{\urlprefix }}%
\providecommand \urlprefix  [0]{URL }%
\providecommand \Eprint [0]{\href }%
\providecommand \doibase [0]{https://doi.org/}%
\providecommand \selectlanguage [0]{\@gobble}%
\providecommand \bibinfo  [0]{\@secondoftwo}%
\providecommand \bibfield  [0]{\@secondoftwo}%
\providecommand \translation [1]{[#1]}%
\providecommand \BibitemOpen [0]{}%
\providecommand \bibitemStop [0]{}%
\providecommand \bibitemNoStop [0]{.\EOS\space}%
\providecommand \EOS [0]{\spacefactor3000\relax}%
\providecommand \BibitemShut  [1]{\csname bibitem#1\endcsname}%
\let\auto@bib@innerbib\@empty
\bibitem [{\citenamefont {Gasser}\ \emph {et~al.}(2008)\citenamefont {Gasser},
  \citenamefont {Lyubovitskij},\ and\ \citenamefont
  {Rusetsky}}]{Gasser:2007zt}%
  \BibitemOpen
  \bibfield  {author} {\bibinfo {author} {\bibfnamefont {J.}~\bibnamefont
  {Gasser}}, \bibinfo {author} {\bibfnamefont {V.~E.}\ \bibnamefont
  {Lyubovitskij}},\ and\ \bibinfo {author} {\bibfnamefont {A.}~\bibnamefont
  {Rusetsky}},\ }\bibfield  {title} {\bibinfo {title} {{Hadronic atoms in QCD +
  QED}},\ }\href {https://doi.org/10.1016/j.physrep.2007.09.006} {\bibfield
  {journal} {\bibinfo  {journal} {Phys. Rept.}\ }\textbf {\bibinfo {volume}
  {456}},\ \bibinfo {pages} {167} (\bibinfo {year} {2008})},\ \Eprint
  {https://arxiv.org/abs/0711.3522} {arXiv:0711.3522 [hep-ph]} \BibitemShut
  {NoStop}%
\bibitem [{\citenamefont {Curceanu}\ \emph {et~al.}(2026)\citenamefont
  {Curceanu}, \citenamefont {Sgaramella}, \citenamefont {Bazzi}, \citenamefont
  {Hashimoto}, \citenamefont {Iliescu}, \citenamefont {Scordo}, \citenamefont
  {Sirghi},\ and\ \citenamefont {Sirghi}}]{Curceanu:2026zjg}%
  \BibitemOpen
  \bibfield  {author} {\bibinfo {author} {\bibfnamefont {C.}~\bibnamefont
  {Curceanu}}, \bibinfo {author} {\bibfnamefont {F.}~\bibnamefont
  {Sgaramella}}, \bibinfo {author} {\bibfnamefont {M.}~\bibnamefont {Bazzi}},
  \bibinfo {author} {\bibfnamefont {T.}~\bibnamefont {Hashimoto}}, \bibinfo
  {author} {\bibfnamefont {M.}~\bibnamefont {Iliescu}}, \bibinfo {author}
  {\bibfnamefont {A.}~\bibnamefont {Scordo}}, \bibinfo {author} {\bibfnamefont
  {D.}~\bibnamefont {Sirghi}},\ and\ \bibinfo {author} {\bibfnamefont
  {F.}~\bibnamefont {Sirghi}},\ }\bibfield  {title} {\bibinfo {title} {{Light
  kaonic atoms as probes of fundamental interactions in strange systems}},\
  }\href {https://doi.org/10.1016/j.ppnp.2026.104226} {\bibfield  {journal}
  {\bibinfo  {journal} {Prog. Part. Nucl. Phys.}\ }\textbf {\bibinfo {volume}
  {147}},\ \bibinfo {pages} {104226} (\bibinfo {year} {2026})}\BibitemShut
  {NoStop}%
\bibitem [{\citenamefont {Dalitz}\ and\ \citenamefont
  {Tuan}(1959)}]{Dalitz:1959dn}%
  \BibitemOpen
  \bibfield  {author} {\bibinfo {author} {\bibfnamefont {R.~H.}\ \bibnamefont
  {Dalitz}}\ and\ \bibinfo {author} {\bibfnamefont {S.~F.}\ \bibnamefont
  {Tuan}},\ }\bibfield  {title} {\bibinfo {title} {{A possible resonant state
  in pion-hyperon scattering}},\ }\href
  {https://doi.org/10.1103/PhysRevLett.2.425} {\bibfield  {journal} {\bibinfo
  {journal} {Phys. Rev. Lett.}\ }\textbf {\bibinfo {volume} {2}},\ \bibinfo
  {pages} {425} (\bibinfo {year} {1959})}\BibitemShut {NoStop}%
\bibitem [{\citenamefont {Hyodo}\ and\ \citenamefont
  {Jido}(2012)}]{Hyodo:2011ur}%
  \BibitemOpen
  \bibfield  {author} {\bibinfo {author} {\bibfnamefont {T.}~\bibnamefont
  {Hyodo}}\ and\ \bibinfo {author} {\bibfnamefont {D.}~\bibnamefont {Jido}},\
  }\bibfield  {title} {\bibinfo {title} {{The nature of the Lambda(1405)
  resonance in chiral dynamics}},\ }\href
  {https://doi.org/10.1016/j.ppnp.2011.07.002} {\bibfield  {journal} {\bibinfo
  {journal} {Prog. Part. Nucl. Phys.}\ }\textbf {\bibinfo {volume} {67}},\
  \bibinfo {pages} {55} (\bibinfo {year} {2012})},\ \Eprint
  {https://arxiv.org/abs/1104.4474} {arXiv:1104.4474 [nucl-th]} \BibitemShut
  {NoStop}%
\bibitem [{\citenamefont {Mai}(2021)}]{Mai:2020ltx}%
  \BibitemOpen
  \bibfield  {author} {\bibinfo {author} {\bibfnamefont {M.}~\bibnamefont
  {Mai}},\ }\bibfield  {title} {\bibinfo {title} {{Review of the ${\Lambda
  }$(1405) A curious case of a strangeness resonance}},\ }\href
  {https://doi.org/10.1140/epjs/s11734-021-00144-7} {\bibfield  {journal}
  {\bibinfo  {journal} {Eur. Phys. J. ST}\ }\textbf {\bibinfo {volume} {230}},\
  \bibinfo {pages} {1593} (\bibinfo {year} {2021})},\ \Eprint
  {https://arxiv.org/abs/2010.00056} {arXiv:2010.00056 [nucl-th]} \BibitemShut
  {NoStop}%
\bibitem [{\citenamefont {Guo}\ \emph {et~al.}(2018)\citenamefont {Guo},
  \citenamefont {Hanhart}, \citenamefont {Mei{\ss}ner}, \citenamefont {Wang},
  \citenamefont {Zhao},\ and\ \citenamefont {Zou}}]{Guo:2017jvc}%
  \BibitemOpen
  \bibfield  {author} {\bibinfo {author} {\bibfnamefont {F.-K.}\ \bibnamefont
  {Guo}}, \bibinfo {author} {\bibfnamefont {C.}~\bibnamefont {Hanhart}},
  \bibinfo {author} {\bibfnamefont {U.-G.}\ \bibnamefont {Mei{\ss}ner}},
  \bibinfo {author} {\bibfnamefont {Q.}~\bibnamefont {Wang}}, \bibinfo {author}
  {\bibfnamefont {Q.}~\bibnamefont {Zhao}},\ and\ \bibinfo {author}
  {\bibfnamefont {B.-S.}\ \bibnamefont {Zou}},\ }\bibfield  {title} {\bibinfo
  {title} {{Hadronic molecules}},\ }\href
  {https://doi.org/10.1103/RevModPhys.90.015004} {\bibfield  {journal}
  {\bibinfo  {journal} {Rev. Mod. Phys.}\ }\textbf {\bibinfo {volume} {90}},\
  \bibinfo {pages} {015004} (\bibinfo {year} {2018})},\ \bibinfo {note}
  {[Erratum: Rev.Mod.Phys. 94, 029901 (2022)]},\ \Eprint
  {https://arxiv.org/abs/1705.00141} {arXiv:1705.00141 [hep-ph]} \BibitemShut
  {NoStop}%
\bibitem [{\citenamefont {Ikeda}\ \emph {et~al.}(2012)\citenamefont {Ikeda},
  \citenamefont {Hyodo},\ and\ \citenamefont {Weise}}]{Ikeda:2012au}%
  \BibitemOpen
  \bibfield  {author} {\bibinfo {author} {\bibfnamefont {Y.}~\bibnamefont
  {Ikeda}}, \bibinfo {author} {\bibfnamefont {T.}~\bibnamefont {Hyodo}},\ and\
  \bibinfo {author} {\bibfnamefont {W.}~\bibnamefont {Weise}},\ }\bibfield
  {title} {\bibinfo {title} {{Chiral SU(3) theory of antikaon-nucleon
  interactions with improved threshold constraints}},\ }\href
  {https://doi.org/10.1016/j.nuclphysa.2012.01.029} {\bibfield  {journal}
  {\bibinfo  {journal} {Nucl. Phys. A}\ }\textbf {\bibinfo {volume} {881}},\
  \bibinfo {pages} {98} (\bibinfo {year} {2012})}\BibitemShut {NoStop}%
\bibitem [{\citenamefont {Guo}\ and\ \citenamefont {Oller}(2013)}]{Guo:2012vv}%
  \BibitemOpen
  \bibfield  {author} {\bibinfo {author} {\bibfnamefont {Z.-H.}\ \bibnamefont
  {Guo}}\ and\ \bibinfo {author} {\bibfnamefont {J.~A.}\ \bibnamefont
  {Oller}},\ }\bibfield  {title} {\bibinfo {title} {{Meson-baryon reactions
  with strangeness -1 within a chiral framework}},\ }\href
  {https://doi.org/10.1103/PhysRevC.87.035202} {\bibfield  {journal} {\bibinfo
  {journal} {Phys. Rev. C}\ }\textbf {\bibinfo {volume} {87}},\ \bibinfo
  {pages} {035202} (\bibinfo {year} {2013})}\BibitemShut {NoStop}%
\bibitem [{\citenamefont {Mai}\ and\ \citenamefont
  {Mei\ss{}ner}(2015)}]{Mai:2014xna}%
  \BibitemOpen
  \bibfield  {author} {\bibinfo {author} {\bibfnamefont {M.}~\bibnamefont
  {Mai}}\ and\ \bibinfo {author} {\bibfnamefont {U.-G.}\ \bibnamefont
  {Mei\ss{}ner}},\ }\bibfield  {title} {\bibinfo {title} {{Constraints on the
  chiral unitary $\bar KN$ amplitude from $\pi\Sigma K^+$ photoproduction
  data}},\ }\href {https://doi.org/10.1140/epja/i2015-15030-3} {\bibfield
  {journal} {\bibinfo  {journal} {Eur. Phys. J. A}\ }\textbf {\bibinfo {volume}
  {51}},\ \bibinfo {pages} {30} (\bibinfo {year} {2015})}\BibitemShut {NoStop}%
\bibitem [{\citenamefont {Cieply}\ and\ \citenamefont
  {Smejkal}(2012)}]{Cieply:2011nq}%
  \BibitemOpen
  \bibfield  {author} {\bibinfo {author} {\bibfnamefont {A.}~\bibnamefont
  {Cieply}}\ and\ \bibinfo {author} {\bibfnamefont {J.}~\bibnamefont
  {Smejkal}},\ }\bibfield  {title} {\bibinfo {title} {{Chirally motivated
  $\bar{K}N$ amplitudes for in-medium applications}},\ }\href
  {https://doi.org/10.1016/j.nuclphysa.2012.01.028} {\bibfield  {journal}
  {\bibinfo  {journal} {Nucl. Phys. A}\ }\textbf {\bibinfo {volume} {881}},\
  \bibinfo {pages} {115} (\bibinfo {year} {2012})}\BibitemShut {NoStop}%
\bibitem [{\citenamefont {Feijoo}\ \emph {et~al.}(2019)\citenamefont {Feijoo},
  \citenamefont {Magas},\ and\ \citenamefont {Ramos}}]{Feijoo:2018den}%
  \BibitemOpen
  \bibfield  {author} {\bibinfo {author} {\bibfnamefont {A.}~\bibnamefont
  {Feijoo}}, \bibinfo {author} {\bibfnamefont {V.}~\bibnamefont {Magas}},\ and\
  \bibinfo {author} {\bibfnamefont {A.}~\bibnamefont {Ramos}},\ }\bibfield
  {title} {\bibinfo {title} {{$S$=\ensuremath{-}1 meson-baryon interaction and
  the role of isospin filtering processes}},\ }\href
  {https://doi.org/10.1103/PhysRevC.99.035211} {\bibfield  {journal} {\bibinfo
  {journal} {Phys. Rev. C}\ }\textbf {\bibinfo {volume} {99}},\ \bibinfo
  {pages} {035211} (\bibinfo {year} {2019})}\BibitemShut {NoStop}%
\bibitem [{\citenamefont {Bazzi}\ \emph {et~al.}(2011)\citenamefont {Bazzi}
  \emph {et~al.}}]{SIDDHARTA:2011dsy}%
  \BibitemOpen
  \bibfield  {author} {\bibinfo {author} {\bibfnamefont {M.}~\bibnamefont
  {Bazzi}} \emph {et~al.} (\bibinfo {collaboration} {SIDDHARTA}),\ }\bibfield
  {title} {\bibinfo {title} {{A New Measurement of Kaonic Hydrogen X-rays}},\
  }\href {https://doi.org/10.1016/j.physletb.2011.09.011} {\bibfield  {journal}
  {\bibinfo  {journal} {Phys. Lett. B}\ }\textbf {\bibinfo {volume} {704}},\
  \bibinfo {pages} {113} (\bibinfo {year} {2011})}\BibitemShut {NoStop}%
\bibitem [{\citenamefont {Kamalov}\ \emph {et~al.}(2001)\citenamefont
  {Kamalov}, \citenamefont {Oset},\ and\ \citenamefont
  {Ramos}}]{Kamalov:2000iy}%
  \BibitemOpen
  \bibfield  {author} {\bibinfo {author} {\bibfnamefont {S.~S.}\ \bibnamefont
  {Kamalov}}, \bibinfo {author} {\bibfnamefont {E.}~\bibnamefont {Oset}},\ and\
  \bibinfo {author} {\bibfnamefont {A.}~\bibnamefont {Ramos}},\ }\bibfield
  {title} {\bibinfo {title} {{Chiral unitary approach to the K- deuteron
  scattering length}},\ }\href {https://doi.org/10.1016/S0375-9474(00)00709-0}
  {\bibfield  {journal} {\bibinfo  {journal} {Nucl. Phys. A}\ }\textbf
  {\bibinfo {volume} {690}},\ \bibinfo {pages} {494} (\bibinfo {year}
  {2001})}\BibitemShut {NoStop}%
\bibitem [{\citenamefont {Ramos}\ \emph {et~al.}(2026)\citenamefont {Ramos},
  \citenamefont {Torres-Rincon}, \citenamefont {de~Fagoaga},\ and\
  \citenamefont {Cabr{\'e}}}]{Ramos:2025ibe}%
  \BibitemOpen
  \bibfield  {author} {\bibinfo {author} {\bibfnamefont {{\`A}.}~\bibnamefont
  {Ramos}}, \bibinfo {author} {\bibfnamefont {J.~M.}\ \bibnamefont
  {Torres-Rincon}}, \bibinfo {author} {\bibfnamefont {A.}~\bibnamefont
  {de~Fagoaga}},\ and\ \bibinfo {author} {\bibfnamefont {E.}~\bibnamefont
  {Cabr{\'e}}},\ }\bibfield  {title} {\bibinfo {title} {{Kaon-deuteron
  femtoscopy from unitarized chiral interactions}},\ }\href
  {https://doi.org/10.1103/bhn4-m9nf} {\bibfield  {journal} {\bibinfo
  {journal} {Phys. Rev. D}\ }\textbf {\bibinfo {volume} {113}},\ \bibinfo
  {pages} {036020} (\bibinfo {year} {2026})},\ \Eprint
  {https://arxiv.org/abs/2507.22593} {arXiv:2507.22593 [hep-ph]} \BibitemShut
  {NoStop}%
\bibitem [{\citenamefont {R\'evai}(2016)}]{Revai:2016muw}%
  \BibitemOpen
  \bibfield  {author} {\bibinfo {author} {\bibfnamefont {J.}~\bibnamefont
  {R\'evai}},\ }\bibfield  {title} {\bibinfo {title} {{Three-body calculation
  of the $1s$ level shift in kaonic deuterium with realistic $\bar{K}N$
  potentials}},\ }\href {https://doi.org/10.1103/PhysRevC.94.054001} {\bibfield
   {journal} {\bibinfo  {journal} {Phys. Rev. C}\ }\textbf {\bibinfo {volume}
  {94}},\ \bibinfo {pages} {054001} (\bibinfo {year} {2016})}\BibitemShut
  {NoStop}%
\bibitem [{\citenamefont {Shevchenko}(2012)}]{Shevchenko:2011ce}%
  \BibitemOpen
  \bibfield  {author} {\bibinfo {author} {\bibfnamefont {N.~V.}\ \bibnamefont
  {Shevchenko}},\ }\bibfield  {title} {\bibinfo {title} {{One- versus two-pole
  $\bar{K}N - \pi \Sigma$ potential: $K^- d$ scattering length}},\ }\href
  {https://doi.org/10.1103/PhysRevC.85.034001} {\bibfield  {journal} {\bibinfo
  {journal} {Phys. Rev. C}\ }\textbf {\bibinfo {volume} {85}},\ \bibinfo
  {pages} {034001} (\bibinfo {year} {2012})}\BibitemShut {NoStop}%
\bibitem [{\citenamefont {Mizutani}\ \emph {et~al.}(2013)\citenamefont
  {Mizutani}, \citenamefont {Fayard}, \citenamefont {Saghai},\ and\
  \citenamefont {Tsushima}}]{Mizutani:2012gy}%
  \BibitemOpen
  \bibfield  {author} {\bibinfo {author} {\bibfnamefont {T.}~\bibnamefont
  {Mizutani}}, \bibinfo {author} {\bibfnamefont {C.}~\bibnamefont {Fayard}},
  \bibinfo {author} {\bibfnamefont {B.}~\bibnamefont {Saghai}},\ and\ \bibinfo
  {author} {\bibfnamefont {K.}~\bibnamefont {Tsushima}},\ }\bibfield  {title}
  {\bibinfo {title} {{Faddeev-chiral unitary approach to the K\ensuremath{-}d
  scattering length}},\ }\href {https://doi.org/10.1103/PhysRevC.87.035201}
  {\bibfield  {journal} {\bibinfo  {journal} {Phys. Rev. C}\ }\textbf {\bibinfo
  {volume} {87}},\ \bibinfo {pages} {035201} (\bibinfo {year}
  {2013})}\BibitemShut {NoStop}%
\bibitem [{\citenamefont {Friedman}\ and\ \citenamefont
  {Gal}(2012)}]{Friedman:2012pc}%
  \BibitemOpen
  \bibfield  {author} {\bibinfo {author} {\bibfnamefont {E.}~\bibnamefont
  {Friedman}}\ and\ \bibinfo {author} {\bibfnamefont {A.}~\bibnamefont {Gal}},\
  }\bibfield  {title} {\bibinfo {title} {{Kaonic atoms and in-medium $K^-N$
  amplitudes}},\ }\href {https://doi.org/10.1016/j.nuclphysa.2012.01.005}
  {\bibfield  {journal} {\bibinfo  {journal} {Nucl. Phys. A}\ }\textbf
  {\bibinfo {volume} {881}},\ \bibinfo {pages} {150} (\bibinfo {year}
  {2012})},\ \Eprint {https://arxiv.org/abs/1201.3770} {arXiv:1201.3770
  [nucl-th]} \BibitemShut {NoStop}%
\bibitem [{\citenamefont {De~Pietri}\ \emph {et~al.}(2019)\citenamefont
  {De~Pietri}, \citenamefont {Drago}, \citenamefont {Feo}, \citenamefont
  {Pagliara}, \citenamefont {Pasquali}, \citenamefont {Traversi},\ and\
  \citenamefont {Wiktorowicz}}]{DePietri:2019khb}%
  \BibitemOpen
  \bibfield  {author} {\bibinfo {author} {\bibfnamefont {R.}~\bibnamefont
  {De~Pietri}}, \bibinfo {author} {\bibfnamefont {A.}~\bibnamefont {Drago}},
  \bibinfo {author} {\bibfnamefont {A.}~\bibnamefont {Feo}}, \bibinfo {author}
  {\bibfnamefont {G.}~\bibnamefont {Pagliara}}, \bibinfo {author}
  {\bibfnamefont {M.}~\bibnamefont {Pasquali}}, \bibinfo {author}
  {\bibfnamefont {S.}~\bibnamefont {Traversi}},\ and\ \bibinfo {author}
  {\bibfnamefont {G.}~\bibnamefont {Wiktorowicz}},\ }\bibfield  {title}
  {\bibinfo {title} {{Merger of compact stars in the two-families scenario}},\
  }\href {https://doi.org/10.3847/1538-4357/ab2fd0} {\bibfield  {journal}
  {\bibinfo  {journal} {Astrophys. J.}\ }\textbf {\bibinfo {volume} {881}},\
  \bibinfo {pages} {122} (\bibinfo {year} {2019})}\BibitemShut {NoStop}%
\bibitem [{\citenamefont {Tolos}\ and\ \citenamefont
  {Fabbietti}(2020)}]{Tolos:2020aln}%
  \BibitemOpen
  \bibfield  {author} {\bibinfo {author} {\bibfnamefont {L.}~\bibnamefont
  {Tolos}}\ and\ \bibinfo {author} {\bibfnamefont {L.}~\bibnamefont
  {Fabbietti}},\ }\bibfield  {title} {\bibinfo {title} {{Strangeness in Nuclei
  and Neutron Stars}},\ }\href {https://doi.org/10.1016/j.ppnp.2020.103770}
  {\bibfield  {journal} {\bibinfo  {journal} {Prog. Part. Nucl. Phys.}\
  }\textbf {\bibinfo {volume} {112}},\ \bibinfo {pages} {103770} (\bibinfo
  {year} {2020})}\BibitemShut {NoStop}%
\bibitem [{\citenamefont {Merafina}\ \emph {et~al.}(2020)\citenamefont
  {Merafina}, \citenamefont {Saturni}, \citenamefont {Curceanu}, \citenamefont
  {Del~Grande},\ and\ \citenamefont {Piscicchia}}]{Merafina:2020ffb}%
  \BibitemOpen
  \bibfield  {author} {\bibinfo {author} {\bibfnamefont {M.}~\bibnamefont
  {Merafina}}, \bibinfo {author} {\bibfnamefont {F.~G.}\ \bibnamefont
  {Saturni}}, \bibinfo {author} {\bibfnamefont {C.}~\bibnamefont {Curceanu}},
  \bibinfo {author} {\bibfnamefont {R.}~\bibnamefont {Del~Grande}},\ and\
  \bibinfo {author} {\bibfnamefont {K.}~\bibnamefont {Piscicchia}},\ }\bibfield
   {title} {\bibinfo {title} {{Self-gravitating strange dark matter halos
  around galaxies}},\ }\href {https://doi.org/10.1103/PhysRevD.102.083015}
  {\bibfield  {journal} {\bibinfo  {journal} {Phys. Rev. D}\ }\textbf {\bibinfo
  {volume} {102}},\ \bibinfo {pages} {083015} (\bibinfo {year}
  {2020})}\BibitemShut {NoStop}%
\bibitem [{\citenamefont {Mishra}\ \emph {et~al.}(2010)\citenamefont {Mishra},
  \citenamefont {Kumar}, \citenamefont {Sanyal}, \citenamefont {Dexheimer},\
  and\ \citenamefont {Schramm}}]{Mishra:2010Kaons}%
  \BibitemOpen
  \bibfield  {author} {\bibinfo {author} {\bibfnamefont {A.}~\bibnamefont
  {Mishra}}, \bibinfo {author} {\bibfnamefont {A.}~\bibnamefont {Kumar}},
  \bibinfo {author} {\bibfnamefont {S.}~\bibnamefont {Sanyal}}, \bibinfo
  {author} {\bibfnamefont {V.}~\bibnamefont {Dexheimer}},\ and\ \bibinfo
  {author} {\bibfnamefont {S.}~\bibnamefont {Schramm}},\ }\bibfield  {title}
  {\bibinfo {title} {{Kaon properties in (proto-)neutron star matter}},\ }\href
  {https://doi.org/10.1140/epja/i2010-10986-x} {\bibfield  {journal} {\bibinfo
  {journal} {Eur. Phys. J. A}\ }\textbf {\bibinfo {volume} {45}},\ \bibinfo
  {pages} {169} (\bibinfo {year} {2010})},\ \Eprint
  {https://arxiv.org/abs/0905.3518} {arXiv:0905.3518 [nucl-th]} \BibitemShut
  {NoStop}%
\bibitem [{\citenamefont {Thapa}\ and\ \citenamefont
  {Sinha}(2020)}]{Thapa:2020Kaons}%
  \BibitemOpen
  \bibfield  {author} {\bibinfo {author} {\bibfnamefont {V.~B.}\ \bibnamefont
  {Thapa}}\ and\ \bibinfo {author} {\bibfnamefont {M.}~\bibnamefont {Sinha}},\
  }\bibfield  {title} {\bibinfo {title} {{Dense matter equation of state of a
  massive neutron star with antikaon condensation}},\ }\href
  {https://doi.org/10.1103/PhysRevD.102.123007} {\bibfield  {journal} {\bibinfo
   {journal} {Phys. Rev. D}\ }\textbf {\bibinfo {volume} {102}},\ \bibinfo
  {pages} {123007} (\bibinfo {year} {2020})},\ \Eprint
  {https://arxiv.org/abs/2011.06440} {arXiv:2011.06440 [nucl-th]} \BibitemShut
  {NoStop}%
\bibitem [{\citenamefont {Yamazaki}(2007)}]{Yamazaki:2007yc}%
  \BibitemOpen
  \bibfield  {author} {\bibinfo {author} {\bibfnamefont {T.}~\bibnamefont
  {Yamazaki}},\ }\bibfield  {title} {\bibinfo {title} {{The Quest for Pionic
  and Kaonic Nuclear Bound Systems Following Yukawa and Tomonaga}},\ }\href
  {https://doi.org/10.1143/PTPS.170.138} {\bibfield  {journal} {\bibinfo
  {journal} {Prog. Theor. Phys. Suppl.}\ }\textbf {\bibinfo {volume} {170}},\
  \bibinfo {pages} {138} (\bibinfo {year} {2007})},\ \Eprint
  {https://arxiv.org/abs/0709.1408} {arXiv:0709.1408 [nucl-ex]} \BibitemShut
  {NoStop}%
\bibitem [{\citenamefont {Akaishi}\ \emph {et~al.}(2003)\citenamefont
  {Akaishi}, \citenamefont {Dote},\ and\ \citenamefont
  {Yamazaki}}]{Akaishi:2003jk}%
  \BibitemOpen
  \bibfield  {author} {\bibinfo {author} {\bibfnamefont {Y.}~\bibnamefont
  {Akaishi}}, \bibinfo {author} {\bibfnamefont {A.}~\bibnamefont {Dote}},\ and\
  \bibinfo {author} {\bibfnamefont {T.}~\bibnamefont {Yamazaki}},\ }\bibfield
  {title} {\bibinfo {title} {{Properties of nuclear anti-K bound states}},\
  }\href {https://doi.org/10.1143/PTPS.149.221} {\bibfield  {journal} {\bibinfo
   {journal} {Prog. Theor. Phys. Suppl.}\ }\textbf {\bibinfo {volume} {149}},\
  \bibinfo {pages} {221} (\bibinfo {year} {2003})}\BibitemShut {NoStop}%
\bibitem [{\citenamefont {Dot{\'e}}\ \emph {et~al.}(2008)\citenamefont
  {Dot{\'e}}, \citenamefont {Hyodo},\ and\ \citenamefont
  {Weise}}]{Dote:2008Kpp}%
  \BibitemOpen
  \bibfield  {author} {\bibinfo {author} {\bibfnamefont {A.}~\bibnamefont
  {Dot{\'e}}}, \bibinfo {author} {\bibfnamefont {T.}~\bibnamefont {Hyodo}},\
  and\ \bibinfo {author} {\bibfnamefont {W.}~\bibnamefont {Weise}},\ }\bibfield
   {title} {\bibinfo {title} {{$K^-pp$ system with chiral SU(3) effective
  interaction}},\ }\href {https://doi.org/10.1016/j.nuclphysa.2008.02.001}
  {\bibfield  {journal} {\bibinfo  {journal} {Nucl. Phys. A}\ }\textbf
  {\bibinfo {volume} {804}},\ \bibinfo {pages} {197} (\bibinfo {year}
  {2008})},\ \Eprint {https://arxiv.org/abs/0802.0238} {arXiv:0802.0238
  [nucl-th]} \BibitemShut {NoStop}%
\bibitem [{\citenamefont {Dot{\'e}}\ \emph {et~al.}(2018)\citenamefont
  {Dot{\'e}}, \citenamefont {Inoue},\ and\ \citenamefont {Myo}}]{Dote:2018Kpp}%
  \BibitemOpen
  \bibfield  {author} {\bibinfo {author} {\bibfnamefont {A.}~\bibnamefont
  {Dot{\'e}}}, \bibinfo {author} {\bibfnamefont {T.}~\bibnamefont {Inoue}},\
  and\ \bibinfo {author} {\bibfnamefont {T.}~\bibnamefont {Myo}},\ }\bibfield
  {title} {\bibinfo {title} {{Fully coupled-channel study of $K^-pp$ resonance
  in a chiral SU(3)-based $\bar{K}N$ potential}},\ }\href
  {https://doi.org/10.1016/j.physletb.2018.08.029} {\bibfield  {journal}
  {\bibinfo  {journal} {Phys. Lett. B}\ }\textbf {\bibinfo {volume} {784}},\
  \bibinfo {pages} {405} (\bibinfo {year} {2018})},\ \Eprint
  {https://arxiv.org/abs/1710.07589} {arXiv:1710.07589 [nucl-th]} \BibitemShut
  {NoStop}%
\bibitem [{\citenamefont {Shevchenko}(2017)}]{Shevchenko:2016wnu}%
  \BibitemOpen
  \bibfield  {author} {\bibinfo {author} {\bibfnamefont {N.~V.}\ \bibnamefont
  {Shevchenko}},\ }\bibfield  {title} {\bibinfo {title} {{Three-Body
  Antikaon{\textendash}Nucleon Systems}},\ }\href
  {https://doi.org/10.1007/s00601-016-1170-5} {\bibfield  {journal} {\bibinfo
  {journal} {Few Body Syst.}\ }\textbf {\bibinfo {volume} {58}},\ \bibinfo
  {pages} {6} (\bibinfo {year} {2017})}\BibitemShut {NoStop}%
\bibitem [{\citenamefont {Bazzi}\ \emph
  {et~al.}(2013{\natexlab{a}})\citenamefont {Bazzi} \emph
  {et~al.}}]{SIDDHARTA:2013ftj}%
  \BibitemOpen
  \bibfield  {author} {\bibinfo {author} {\bibfnamefont {M.}~\bibnamefont
  {Bazzi}} \emph {et~al.} (\bibinfo {collaboration} {SIDDHARTA}),\ }\bibfield
  {title} {\bibinfo {title} {{Preliminary study of kaonic deuterium X-rays by
  the SIDDHARTA experiment at DAFNE}},\ }\href
  {https://doi.org/10.1016/j.nuclphysa.2013.03.001} {\bibfield  {journal}
  {\bibinfo  {journal} {Nucl. Phys. A}\ }\textbf {\bibinfo {volume} {907}},\
  \bibinfo {pages} {69} (\bibinfo {year} {2013}{\natexlab{a}})}\BibitemShut
  {NoStop}%
\bibitem [{\citenamefont {Sirghi}\ \emph {et~al.}(2024)\citenamefont {Sirghi}
  \emph {et~al.}}]{Sirghi:2023wok}%
  \BibitemOpen
  \bibfield  {author} {\bibinfo {author} {\bibfnamefont {F.}~\bibnamefont
  {Sirghi}} \emph {et~al.},\ }\bibfield  {title} {\bibinfo {title}
  {{SIDDHARTA-2 apparatus for kaonic atoms research on the
  DA\ensuremath{\Phi}NE collider}},\ }\href
  {https://doi.org/10.1088/1748-0221/19/11/P11006} {\bibfield  {journal}
  {\bibinfo  {journal} {JINST}\ }\textbf {\bibinfo {volume} {19}}\bibinfo
  {number} { (11)},\ \bibinfo {pages} {P11006}}\BibitemShut {NoStop}%
\bibitem [{\citenamefont {Jensen}\ and\ \citenamefont
  {Markushin}(2002{\natexlab{a}})}]{Jensen:2002wq}%
  \BibitemOpen
\bibfield  {number} {  }\bibfield  {author} {\bibinfo {author} {\bibfnamefont
  {T.~S.}\ \bibnamefont {Jensen}}\ and\ \bibinfo {author} {\bibfnamefont
  {V.~E.}\ \bibnamefont {Markushin}},\ }\bibfield  {title} {\bibinfo {title}
  {{Collisional deexcitation of exotic hydrogen atoms in highly excited states.
  1. Cross-sections}},\ }\href {https://doi.org/10.1140/epjd/e2002-00207-y}
  {\bibfield  {journal} {\bibinfo  {journal} {Eur. Phys. J. D}\ }\textbf
  {\bibinfo {volume} {21}},\ \bibinfo {pages} {261} (\bibinfo {year}
  {2002}{\natexlab{a}})}\BibitemShut {NoStop}%
\bibitem [{\citenamefont {Jensen}\ and\ \citenamefont
  {Markushin}(2002{\natexlab{b}})}]{Jensen:2002wr}%
  \BibitemOpen
  \bibfield  {author} {\bibinfo {author} {\bibfnamefont {T.~S.}\ \bibnamefont
  {Jensen}}\ and\ \bibinfo {author} {\bibfnamefont {V.~E.}\ \bibnamefont
  {Markushin}},\ }\bibfield  {title} {\bibinfo {title} {{Collisional
  deexcitation of exotic hydrogen atoms in highly excited states. 2. Cascade
  calculations}},\ }\href {https://doi.org/10.1140/epjd/e2002-00208-x}
  {\bibfield  {journal} {\bibinfo  {journal} {Eur. Phys. J. D}\ }\textbf
  {\bibinfo {volume} {21}},\ \bibinfo {pages} {271} (\bibinfo {year}
  {2002}{\natexlab{b}})}\BibitemShut {NoStop}%
\bibitem [{\citenamefont {Zobov}\ \emph {et~al.}(2018)\citenamefont {Zobov}
  \emph {et~al.}}]{Zobov:2018yxf}%
  \BibitemOpen
  \bibfield  {author} {\bibinfo {author} {\bibfnamefont {M.}~\bibnamefont
  {Zobov}} \emph {et~al.},\ }\bibfield  {title} {\bibinfo {title} {{DA$\Phi$NE
  Collider with Crab Waist Scheme: from KLOE-2 to SIDDHARTA-2 Experiment}},\
  }\bibfield  {booktitle} {\emph {\bibinfo {booktitle} {{26th Russian Particle
  Accelerator Conference}}},\ }\href
  {https://doi.org/10.18429/JACoW-RuPAC2018-MOXMH01} {\bibfield  {journal}
  {\bibinfo  {journal} {JACoW}\ ,\ \bibinfo {pages} {MOXMH01}} (\bibinfo {year}
  {2018})}\BibitemShut {NoStop}%
\bibitem [{\citenamefont {Zobov}\ \emph {et~al.}(2010)\citenamefont {Zobov}
  \emph {et~al.}}]{Zobov:2010zza}%
  \BibitemOpen
  \bibfield  {author} {\bibinfo {author} {\bibfnamefont {M.}~\bibnamefont
  {Zobov}} \emph {et~al.},\ }\bibfield  {title} {\bibinfo {title} {{Test of
  crab-waist collisions at DAFNE Phi factory}},\ }\href
  {https://doi.org/10.1103/PhysRevLett.104.174801} {\bibfield  {journal}
  {\bibinfo  {journal} {Phys. Rev. Lett.}\ }\textbf {\bibinfo {volume} {104}},\
  \bibinfo {pages} {174801} (\bibinfo {year} {2010})}\BibitemShut {NoStop}%
\bibitem [{\citenamefont {Milardi}\ \emph {et~al.}(2018)\citenamefont {Milardi}
  \emph {et~al.}}]{Milardi:2018sih}%
  \BibitemOpen
  \bibfield  {author} {\bibinfo {author} {\bibfnamefont {C.}~\bibnamefont
  {Milardi}} \emph {et~al.},\ }\bibfield  {title} {\bibinfo {title}
  {{Preparation Activity for the Siddharta-2 Run at DA\ensuremath{\Phi}NE}},\
  }\bibfield  {booktitle} {\emph {\bibinfo {booktitle} {{9th International
  Particle Accelerator Conference, IPAC2018, Vancouver BC Canada}}},\
  }\bibfield  {journal} {\bibinfo  {journal} {JACoW}\ }\href
  {https://doi.org/10.18429/JACoW-IPAC2018-MOPMF088}
  {10.18429/JACoW-IPAC2018-MOPMF088} (\bibinfo {year} {2018})\BibitemShut
  {NoStop}%
\bibitem [{\citenamefont {Milardi}\ \emph {et~al.}(2021)\citenamefont {Milardi}
  \emph {et~al.}}]{Milardi:2021khj}%
  \BibitemOpen
  \bibfield  {author} {\bibinfo {author} {\bibfnamefont {C.}~\bibnamefont
  {Milardi}} \emph {et~al.},\ }\bibfield  {title} {\bibinfo {title}
  {{DA{\ensuremath{\Phi}}NE Commissioning for SIDDHARTA-2 Experiment}},\ }\href
  {https://doi.org/10.18429/JACoW-IPAC2021-TUPAB001} {\bibfield  {journal}
  {\bibinfo  {journal} {JACoW}\ }\textbf {\bibinfo {volume} {IPAC2021}},\
  \bibinfo {pages} {TUPAB001} (\bibinfo {year} {2021})}\BibitemShut {NoStop}%
\bibitem [{\citenamefont {Milardi}\ \emph {et~al.}(2024)\citenamefont {Milardi}
  \emph {et~al.}}]{Milardi:2024efr}%
  \BibitemOpen
  \bibfield  {author} {\bibinfo {author} {\bibfnamefont {C.}~\bibnamefont
  {Milardi}} \emph {et~al.},\ }\bibfield  {title} {\bibinfo {title} {{DAFNE
  operation strategy for the observation of the kaonic deuterium}},\ }\href
  {https://doi.org/10.18429/JACoW-IPAC2024-WEPR17} {\bibfield  {journal}
  {\bibinfo  {journal} {JACoW}\ }\textbf {\bibinfo {volume} {IPAC2024}},\
  \bibinfo {pages} {WEPR17} (\bibinfo {year} {2024})}\BibitemShut {NoStop}%
\bibitem [{\citenamefont {Miliucci}\ \emph {et~al.}(2021)\citenamefont
  {Miliucci} \emph {et~al.}}]{Miliucci:2021wbj}%
  \BibitemOpen
  \bibfield  {author} {\bibinfo {author} {\bibfnamefont {M.}~\bibnamefont
  {Miliucci}} \emph {et~al.},\ }\bibfield  {title} {\bibinfo {title} {{Silicon
  drift detectors system for high-precision light kaonic atoms spectroscopy}},\
  }\href {https://doi.org/10.1088/1361-6501/abeea9} {\bibfield  {journal}
  {\bibinfo  {journal} {Measur. Sci. Tech.}\ }\textbf {\bibinfo {volume}
  {32}},\ \bibinfo {pages} {095501} (\bibinfo {year} {2021})}\BibitemShut
  {NoStop}%
\bibitem [{\citenamefont {Miliucci}\ \emph {et~al.}(2022)\citenamefont
  {Miliucci} \emph {et~al.}}]{Miliucci:2022lvn}%
  \BibitemOpen
  \bibfield  {author} {\bibinfo {author} {\bibfnamefont {M.}~\bibnamefont
  {Miliucci}} \emph {et~al.},\ }\bibfield  {title} {\bibinfo {title} {{Large
  area silicon drift detectors system for high precision timed x-ray
  spectroscopy}},\ }\href {https://doi.org/10.1088/1361-6501/ac777a} {\bibfield
   {journal} {\bibinfo  {journal} {Measur. Sci. Tech.}\ }\textbf {\bibinfo
  {volume} {33}},\ \bibinfo {pages} {095502} (\bibinfo {year}
  {2022})}\BibitemShut {NoStop}%
\bibitem [{\citenamefont {Santos}\ \emph {et~al.}(2005)\citenamefont {Santos},
  \citenamefont {Parente}, \citenamefont {Boucard}, \citenamefont
  {Indelicato},\ and\ \citenamefont {Desclaux}}]{Santos:2004bw}%
  \BibitemOpen
  \bibfield  {author} {\bibinfo {author} {\bibfnamefont {J.~P.}\ \bibnamefont
  {Santos}}, \bibinfo {author} {\bibfnamefont {F.}~\bibnamefont {Parente}},
  \bibinfo {author} {\bibfnamefont {S.}~\bibnamefont {Boucard}}, \bibinfo
  {author} {\bibfnamefont {P.}~\bibnamefont {Indelicato}},\ and\ \bibinfo
  {author} {\bibfnamefont {J.~P.}\ \bibnamefont {Desclaux}},\ }\bibfield
  {title} {\bibinfo {title} {{X-ray energies of circular transitions and
  electrons screening in kaonic atoms}},\ }\href
  {https://doi.org/10.1103/PhysRevA.71.032501} {\bibfield  {journal} {\bibinfo
  {journal} {Phys. Rev. A}\ }\textbf {\bibinfo {volume} {71}},\ \bibinfo
  {pages} {032501} (\bibinfo {year} {2005})}\BibitemShut {NoStop}%
\bibitem [{\citenamefont {Karshenboim}\ \emph
  {et~al.}(2006{\natexlab{a}})\citenamefont {Karshenboim}, \citenamefont
  {Ivanov},\ and\ \citenamefont {Korzinin}}]{Karshenboim:2006zz}%
  \BibitemOpen
  \bibfield  {author} {\bibinfo {author} {\bibfnamefont {S.~G.}\ \bibnamefont
  {Karshenboim}}, \bibinfo {author} {\bibfnamefont {V.~G.}\ \bibnamefont
  {Ivanov}},\ and\ \bibinfo {author} {\bibfnamefont {E.~Y.}\ \bibnamefont
  {Korzinin}},\ }\bibfield  {title} {\bibinfo {title} {{Vacuum polarization in
  muonic atoms: The Lamb shift at low and medium Z}},\ }\href
  {https://doi.org/10.1140/epjd/e2006-00133-0} {\bibfield  {journal} {\bibinfo
  {journal} {Eur. Phys. J. D}\ }\textbf {\bibinfo {volume} {39}},\ \bibinfo
  {pages} {351} (\bibinfo {year} {2006}{\natexlab{a}})},\ \Eprint
  {https://arxiv.org/abs/physics/0604133} {arXiv:physics/0604133} \BibitemShut
  {NoStop}%
\bibitem [{\citenamefont {Karshenboim}\ \emph
  {et~al.}(2006{\natexlab{b}})\citenamefont {Karshenboim}, \citenamefont
  {Korzinin},\ and\ \citenamefont {Ivanov}}]{Karshenboim:2005am}%
  \BibitemOpen
  \bibfield  {author} {\bibinfo {author} {\bibfnamefont {S.~G.}\ \bibnamefont
  {Karshenboim}}, \bibinfo {author} {\bibfnamefont {E.~Y.}\ \bibnamefont
  {Korzinin}},\ and\ \bibinfo {author} {\bibfnamefont {V.~G.}\ \bibnamefont
  {Ivanov}},\ }\bibfield  {title} {\bibinfo {title} {{The Uehling correction to
  the energy levels in a pionic atom}},\ }\href
  {https://doi.org/10.1139/P06-010} {\bibfield  {journal} {\bibinfo  {journal}
  {Can. J. Phys.}\ }\textbf {\bibinfo {volume} {84}},\ \bibinfo {pages} {107}
  (\bibinfo {year} {2006}{\natexlab{b}})},\ \Eprint
  {https://arxiv.org/abs/physics/0510079} {arXiv:physics/0510079} \BibitemShut
  {NoStop}%
\bibitem [{\citenamefont {Gal}(2007)}]{Gal:2006cw}%
  \BibitemOpen
  \bibfield  {author} {\bibinfo {author} {\bibfnamefont {A.}~\bibnamefont
  {Gal}},\ }\bibfield  {title} {\bibinfo {title} {{On the scattering length of
  the K- d system}},\ }\href {https://doi.org/10.1142/S0217751X07035379}
  {\bibfield  {journal} {\bibinfo  {journal} {Int. J. Mod. Phys. A}\ }\textbf
  {\bibinfo {volume} {22}},\ \bibinfo {pages} {226} (\bibinfo {year}
  {2007})}\BibitemShut {NoStop}%
\bibitem [{\citenamefont {Liu}\ \emph {et~al.}(2020)\citenamefont {Liu},
  \citenamefont {Wu}, \citenamefont {Leinweber},\ and\ \citenamefont
  {Thomas}}]{Liu:2020foc}%
  \BibitemOpen
  \bibfield  {author} {\bibinfo {author} {\bibfnamefont {Z.-W.}\ \bibnamefont
  {Liu}}, \bibinfo {author} {\bibfnamefont {J.-J.}\ \bibnamefont {Wu}},
  \bibinfo {author} {\bibfnamefont {D.~B.}\ \bibnamefont {Leinweber}},\ and\
  \bibinfo {author} {\bibfnamefont {A.~W.}\ \bibnamefont {Thomas}},\ }\bibfield
   {title} {\bibinfo {title} {{Kaonic Hydrogen and Deuterium in Hamiltonian
  Effective Field Theory}},\ }\href
  {https://doi.org/10.1016/j.physletb.2020.135652} {\bibfield  {journal}
  {\bibinfo  {journal} {Phys. Lett. B}\ }\textbf {\bibinfo {volume} {808}},\
  \bibinfo {pages} {135652} (\bibinfo {year} {2020})}\BibitemShut {NoStop}%
\bibitem [{\citenamefont {D\"oring}\ and\ \citenamefont
  {Mei\ss{}ner}(2011)}]{Doring:2011xc}%
  \BibitemOpen
  \bibfield  {author} {\bibinfo {author} {\bibfnamefont {M.}~\bibnamefont
  {D\"oring}}\ and\ \bibinfo {author} {\bibfnamefont {U.~G.}\ \bibnamefont
  {Mei\ss{}ner}},\ }\bibfield  {title} {\bibinfo {title}
  {{Kaon\textendash{}nucleon scattering lengths from kaonic deuterium
  experiments revisited}},\ }\href
  {https://doi.org/10.1016/j.physletb.2011.09.099} {\bibfield  {journal}
  {\bibinfo  {journal} {Phys. Lett. B}\ }\textbf {\bibinfo {volume} {704}},\
  \bibinfo {pages} {663} (\bibinfo {year} {2011})}\BibitemShut {NoStop}%
\bibitem [{\citenamefont {Hoshino}\ \emph {et~al.}(2017)\citenamefont
  {Hoshino}, \citenamefont {Ohnishi}, \citenamefont {Horiuchi}, \citenamefont
  {Hyodo},\ and\ \citenamefont {Weise}}]{Hoshino:2017mty}%
  \BibitemOpen
  \bibfield  {author} {\bibinfo {author} {\bibfnamefont {T.}~\bibnamefont
  {Hoshino}}, \bibinfo {author} {\bibfnamefont {S.}~\bibnamefont {Ohnishi}},
  \bibinfo {author} {\bibfnamefont {W.}~\bibnamefont {Horiuchi}}, \bibinfo
  {author} {\bibfnamefont {T.}~\bibnamefont {Hyodo}},\ and\ \bibinfo {author}
  {\bibfnamefont {W.}~\bibnamefont {Weise}},\ }\bibfield  {title} {\bibinfo
  {title} {{Constraining the $\bar{K}N$ interaction from the $1S$ level shift
  of kaonic deuterium}},\ }\href {https://doi.org/10.1103/PhysRevC.96.045204}
  {\bibfield  {journal} {\bibinfo  {journal} {Phys. Rev. C}\ }\textbf {\bibinfo
  {volume} {96}},\ \bibinfo {pages} {045204} (\bibinfo {year}
  {2017})}\BibitemShut {NoStop}%
\bibitem [{\citenamefont {Shevchenko}(2022)}]{Shevchenko:2021swf}%
  \BibitemOpen
  \bibfield  {author} {\bibinfo {author} {\bibfnamefont {N.~V.}\ \bibnamefont
  {Shevchenko}},\ }\bibfield  {title} {\bibinfo {title} {{Light Kaonic Atoms:
  From \textquotedblleft{}Corrected\textquotedblright{} to
  \textquotedblleft{}Summed Up\textquotedblright{} Deser Formula}},\ }\href
  {https://doi.org/10.1007/s00601-022-01727-3} {\bibfield  {journal} {\bibinfo
  {journal} {Few Body Syst.}\ }\textbf {\bibinfo {volume} {63}},\ \bibinfo
  {pages} {22} (\bibinfo {year} {2022})}\BibitemShut {NoStop}%
\bibitem [{\citenamefont {al.}(2026)}]{ALICE:2026pxr}%
  \BibitemOpen
  \bibfield  {author} {\bibinfo {author} {\bibfnamefont {D.~A. H. A.~e.}\
  \bibnamefont {al.}},\ }\bibfield  {title} {\bibinfo {title} {First
  measurement of the strong interaction scattering parameters for the
  ${K}^{\ensuremath{-}}d$ and ${K}^{+}d$ systems},\ }\href
  {https://doi.org/10.1103/5w12-tbm8} {\bibfield  {journal} {\bibinfo
  {journal} {Phys. Rev. Lett.}\ } (\bibinfo {year} {2026})}\BibitemShut
  {NoStop}%
\bibitem [{\citenamefont {Duerinck}\ \emph {et~al.}(2025)\citenamefont
  {Duerinck}, \citenamefont {Lazauskas},\ and\ \citenamefont
  {Dohet-Eraly}}]{Duerinck:2025pbj}%
  \BibitemOpen
  \bibfield  {author} {\bibinfo {author} {\bibfnamefont {P.-Y.}\ \bibnamefont
  {Duerinck}}, \bibinfo {author} {\bibfnamefont {R.}~\bibnamefont
  {Lazauskas}},\ and\ \bibinfo {author} {\bibfnamefont {J.}~\bibnamefont
  {Dohet-Eraly}},\ }\bibfield  {title} {\bibinfo {title} {{Level shifts of
  exotic deuterium atoms from effective range parameters}},\ }\href
  {https://doi.org/10.1103/PhysRevC.111.034003} {\bibfield  {journal} {\bibinfo
   {journal} {Phys. Rev. C}\ }\textbf {\bibinfo {volume} {111}},\ \bibinfo
  {pages} {034003} (\bibinfo {year} {2025})},\ \Eprint
  {https://arxiv.org/abs/2501.14407} {arXiv:2501.14407 [nucl-th]} \BibitemShut
  {NoStop}%
\bibitem [{\citenamefont {Ciepl\'y}\ \emph {et~al.}(2016)\citenamefont
  {Ciepl\'y}, \citenamefont {Mai}, \citenamefont {Mei\ss{}ner},\ and\
  \citenamefont {Smejkal}}]{Cieply:2016jby}%
  \BibitemOpen
  \bibfield  {author} {\bibinfo {author} {\bibfnamefont {A.}~\bibnamefont
  {Ciepl\'y}}, \bibinfo {author} {\bibfnamefont {M.}~\bibnamefont {Mai}},
  \bibinfo {author} {\bibfnamefont {U.-G.}\ \bibnamefont {Mei\ss{}ner}},\ and\
  \bibinfo {author} {\bibfnamefont {J.}~\bibnamefont {Smejkal}},\ }\bibfield
  {title} {\bibinfo {title} {{On the pole content of coupled channels chiral
  approaches used for the $\bar{K}N$ system}},\ }\href
  {https://doi.org/10.1016/j.nuclphysa.2016.04.031} {\bibfield  {journal}
  {\bibinfo  {journal} {Nucl. Phys. A}\ }\textbf {\bibinfo {volume} {954}},\
  \bibinfo {pages} {17} (\bibinfo {year} {2016})}\BibitemShut {NoStop}%
\bibitem [{\citenamefont {Wang}\ \emph {et~al.}(2026)\citenamefont {Wang},
  \citenamefont {Kumar}, \citenamefont {Grefa}, \citenamefont {Maslov},
  \citenamefont {Ratti}, \citenamefont {Negreiros},\ and\ \citenamefont
  {Dexheimer}}]{Wang:2026Kaons}%
  \BibitemOpen
  \bibfield  {author} {\bibinfo {author} {\bibfnamefont {Y.}~\bibnamefont
  {Wang}}, \bibinfo {author} {\bibfnamefont {R.}~\bibnamefont {Kumar}},
  \bibinfo {author} {\bibfnamefont {J.}~\bibnamefont {Grefa}}, \bibinfo
  {author} {\bibfnamefont {K.}~\bibnamefont {Maslov}}, \bibinfo {author}
  {\bibfnamefont {C.}~\bibnamefont {Ratti}}, \bibinfo {author} {\bibfnamefont
  {R.}~\bibnamefont {Negreiros}},\ and\ \bibinfo {author} {\bibfnamefont
  {V.}~\bibnamefont {Dexheimer}},\ }\bibfield  {title} {\bibinfo {title}
  {{$S$-wave kaon condensation in neutron-star matter within a chiral model
  framework with dynamical meson masses}},\ }\href@noop {} {\  (\bibinfo {year}
  {2026})},\ \Eprint {https://arxiv.org/abs/2603.22164} {arXiv:2603.22164
  [nucl-th]} \BibitemShut {NoStop}%
\bibitem [{Sup()}]{SupplementalMaterial}%
  \BibitemOpen
  \href@noop {} {}\bibinfo {note} {See Supplemental Material at [URL will be
  inserted by publisher] for the kaon-trigger time-of-flight distribution, the
  SDD timing distribution and event-selection windows, and the
  electromagnetic-background energy spectrum.}\BibitemShut {Stop}%
\bibitem [{\citenamefont {Bazzi}\ \emph
  {et~al.}(2013{\natexlab{b}})\citenamefont {Bazzi} \emph
  {et~al.}}]{Bazzi:2013kwa}%
  \BibitemOpen
  \bibfield  {author} {\bibinfo {author} {\bibfnamefont {M.}~\bibnamefont
  {Bazzi}} \emph {et~al.},\ }\bibfield  {title} {\bibinfo {title}
  {{Characterization of the SIDDHARTA-2 second level trigger detector prototype
  based on scintillators coupled to a prism reflector light guide}},\ }\href
  {https://doi.org/10.1088/1748-0221/8/11/T11003} {\bibfield  {journal}
  {\bibinfo  {journal} {JINST}\ }\textbf {\bibinfo {volume} {8}},\ \bibinfo
  {pages} {T11003}}\BibitemShut {NoStop}%
\bibitem [{\citenamefont {T\"uchler}\ \emph {et~al.}(2023)\citenamefont
  {T\"uchler} \emph {et~al.}}]{Tuchler:2023ozy}%
  \BibitemOpen
  \bibfield  {author} {\bibinfo {author} {\bibfnamefont {M.}~\bibnamefont
  {T\"uchler}} \emph {et~al.},\ }\bibfield  {title} {\bibinfo {title} {{The
  SIDDHARTA-2 Veto-2 system for X-ray spectroscopy of kaonic atoms at
  DA\ensuremath{\Phi}NE}},\ }\href
  {https://doi.org/10.1088/1748-0221/18/11/P11026} {\bibfield  {journal}
  {\bibinfo  {journal} {JINST}\ }\textbf {\bibinfo {volume} {18}}\bibinfo
  {number} { (11)},\ \bibinfo {pages} {P11026}}\BibitemShut {NoStop}%
\bibitem [{\citenamefont {Sgaramella}\ \emph {et~al.}(2022)\citenamefont
  {Sgaramella} \emph {et~al.}}]{Sgaramella:2022}%
  \BibitemOpen
\bibfield  {number} {  }\bibfield  {author} {\bibinfo {author} {\bibfnamefont
  {F.}~\bibnamefont {Sgaramella}} \emph {et~al.},\ }\bibfield  {title}
  {\bibinfo {title} {The {SIDDHARTA}-2 calibration method for high precision
  kaonic atoms x-ray spectroscopy measurements},\ }\href
  {https://doi.org/10.1088/1402-4896/ac95da} {\bibfield  {journal} {\bibinfo
  {journal} {Physica Scripta}\ }\textbf {\bibinfo {volume} {97}},\ \bibinfo
  {pages} {114002} (\bibinfo {year} {2022})}\BibitemShut {NoStop}%
\bibitem [{\citenamefont {Baru}\ \emph {et~al.}(2009)\citenamefont {Baru},
  \citenamefont {Epelbaum},\ and\ \citenamefont {Rusetsky}}]{Baru:2009tx}%
  \BibitemOpen
  \bibfield  {author} {\bibinfo {author} {\bibfnamefont {V.}~\bibnamefont
  {Baru}}, \bibinfo {author} {\bibfnamefont {E.}~\bibnamefont {Epelbaum}},\
  and\ \bibinfo {author} {\bibfnamefont {A.}~\bibnamefont {Rusetsky}},\
  }\bibfield  {title} {\bibinfo {title} {{The Role of nucleon recoil in
  low-energy antikaon-deuteron scattering}},\ }\href
  {https://doi.org/10.1140/epja/i2009-10845-y} {\bibfield  {journal} {\bibinfo
  {journal} {Eur. Phys. J. A}\ }\textbf {\bibinfo {volume} {42}},\ \bibinfo
  {pages} {111} (\bibinfo {year} {2009})}\BibitemShut {NoStop}%
\bibitem [{\citenamefont {Shevchenko}\ and\ \citenamefont
  {R{\'e}vai}(2014)}]{Shevchenko:2014uva}%
  \BibitemOpen
  \bibfield  {author} {\bibinfo {author} {\bibfnamefont {N.~V.}\ \bibnamefont
  {Shevchenko}}\ and\ \bibinfo {author} {\bibfnamefont {J.}~\bibnamefont
  {R{\'e}vai}},\ }\bibfield  {title} {\bibinfo {title} {{Faddeev calculations
  of the $\bar{K}NN$ system with chirally-motivated $\bar{K}N$ interaction. I.
  Low-energy $K^- d$ scattering and antikaonic deuterium}},\ }\href
  {https://doi.org/10.1103/PhysRevC.90.034003} {\bibfield  {journal} {\bibinfo
  {journal} {Phys. Rev. C}\ }\textbf {\bibinfo {volume} {90}},\ \bibinfo
  {pages} {034003} (\bibinfo {year} {2014})}\BibitemShut {NoStop}%
\end{thebibliography}%

\clearpage
\onecolumngrid
\begin{center}\textbf{END MATTER}\end{center}

\textbf{Event selection.} The event selection combines the kaon trigger system, the Silicon Drift Detectors (SDD) and the hadronic-veto systems of the SIDDHARTA-2 apparatus. The kaon trigger consists of two fast plastic scintillators placed above and below the interaction region and identifies back-to-back low-momentum $K^+K^-$ pairs through their coincidence, with a time resolution of about 350 ps (FWHM). X-rays are detected by 384 SDDs, consisting of 48 monolithic detection modules, with 8 pixels each, arranged around the cryogenic target. Each pixel has an active area of $8\times8~\mathrm{mm}^2$, and the 450 $\mu$m silicon thickness provides nearly 100\% detection efficiency in the energy range relevant to kaonic deuterium spectroscopy. The SDDs operate at 140 K, where the energy resolution is 160 eV FWHM at 6.4 keV, obtained with an analogue semi-Gaussian shaping filter.
The time resolution of 450 ns FWHM is determined by the electron drift time under these cryogenic conditions \cite{Miliucci:2021wbj,Miliucci:2022lvn}.
X-ray events are selected by requiring a coincidence with a valid trigger. Residual minimum-ionizing particles and accidental triggers are suppressed through a time-of-flight selection with respect to the DA$\Phi$NE radio-frequency signal, retaining only the events associated with kaon pairs as shown in Fig. \ref{fig:kt_tof} of Supplemental Material~\cite{SupplementalMaterial}. Candidate SDD hits are then required to fall within the detector timing acceptance window, thereby selecting X-rays correlated with the trigger from the large asynchronous electromagnetic background generated by beam losses and associated showers (Fig. \ref{fig:dt} of Supplemental Material~\cite{SupplementalMaterial}). This combination of trigger, time-of-flight and SDD timing selections provides the primary suppression of the electromagnetic background.
Hadronic background is reduced with three veto systems. Veto-1 is an outer barrel of plastic scintillators surrounding the vacuum chamber and is designed to distinguish kaons stopped in the gaseous target from kaons absorbed in the surrounding solid materials. The method exploits the different moderation times preceding kaon-nuclear absorption. Following the nuclear absorption of a $K^-$, charged secondary particles, predominantly pions, are produced with high probability and cross the apparatus. Veto-1 records the arrival time of these particles relative to the kaon trigger. Because kaons stopping in the gas undergo several collisions before absorption, they produce delayed Veto-1 signals, whereas kaons absorbed in Kapton windows, support structures and other solid elements generate prompt signals. The Veto-1 timing therefore provides an efficient tag of the kaon stop location, allowing the prompt solid-stop component to be rejected \cite{Bazzi:2013kwa}.\\
Additional suppression is obtained with Veto-2 \cite{Tuchler:2023ozy}, an inner ring of plastic-scintillator tiles mounted behind the SDDs and read out by Silicon Photo-Multipliers (SiPMs). Its purpose is to reject background events induced by charged particles crossing the SDDs. Such particles generate signals in the SDDs either directly, through energy deposition in the silicon, or indirectly, through secondary radiation produced in the surrounding materials. Veto-2 records the passage of these charged particles in the scintillator tiles located behind the SDDs. By exploiting the geometrical correlation between a fired SDD and the corresponding Veto-2 tile, together with the adjacent tiles, events associated with traversing charged particles are rejected.
Finally, the charged-kaon veto consists of a plastic scintillator installed below the lower kaon trigger counter and covered by a Teflon layer used to stop the kaons. Its purpose is to identify the charge of the kaon emitted in the direction opposite to the target. When a $K^-$ stops in the absorber, it is promptly captured by an atom and undergoes kaon-nuclear absorption, producing charged secondary particles within nanoseconds. By contrast, a $K^+$ cannot form a bound state and instead decays with a substantially longer lifetime. The detector records the arrival time of these secondary charged particles relative to the kaon trigger, thereby discriminating events associated with an opposite-side $K^-$ from those compatible with a $K^+$. This allows the rejection of events in which the kaon emitted toward the target is not a $K^-$.\\
The timing selections applied to Veto-1 and to the charged-kaon detector are optimized using Geant4 simulations of the SIDDHARTA-2 apparatus. The simulation starts from $\phi$ production at the interaction point, followed by its decay into charged kaon pairs and by the transport of the kaons through the apparatus until they decay or stop in the gas target forming a kaonic atom. When a kaonic atom is created, X-rays are isotropically generated. The charged particles produced after kaon-nuclear absorption are then transported through the setup, and their interactions with the veto detectors are recorded. These simulations are used to define the timing windows that maximized the signal-to-background ratio for kaonic deuterium events.\\ 
The final kaonic deuterium spectrum is obtained after applying the full selection chain. Altogether, these detectors produce a substantial improvement in the achievable signal-to-background ratio compared to SIDDHARTA. This improvement results from several advances: a factor-of-two improvement in SDD time resolution,corresponding to a factor-of-two increase in the rejection of asynchronous electromagnetic background; upgraded trigger, shielding, DA$\Phi$NE optics and collision scheme, which improves the signal-to-background ratio by about a factor of three \cite{Milardi:2024efr}; and a further factor-of-two suppression of hadronic background provided by the veto systems.\\

\textbf{Background evaluation.}
An independent estimate of the electromagnetic background, mainly arising from electromagnetic showers induced by the Touschek effect and beam--gas interactions, is obtained from the SDD timing distribution relative to the kaon trigger (see Fig. \ref{fig:dt} of Supplemental Material~\cite{SupplementalMaterial}).
X-ray events synchronous with kaon production form a prompt peak that defines the signal time window used to construct the kaonic deuterium spectrum. In contrast, the lateral regions, where the distribution is flat, are dominated by the asynchronous electromagnetic background. By selecting events in these sideband regions, a background-only spectrum is constructed and fitted with exponential and constant functions (see Fig. \ref{fig:dt_energy} of Supplemental Material~\cite{SupplementalMaterial}). The parameters and their uncertainties extracted from this fit are used to constrain the corresponding electromagnetic-background parameters in the final fit to the kaonic deuterium spectrum. No analogous data-driven separation is possible for the hadronic background, which is synchronous with kaon production; accordingly, the hadronic contribution is described in the final fit by an exponential component with freely varying parameters.\\

\textbf{Systematic uncertainties.}
The systematic uncertainties include contributions from the SDD calibration, the description of the kaonic deuterium line shape, the background model, and the stability of the analysis against variations in the event-selection criteria. The calibration component is derived from the accuracy of the energy calibration in the region of interest \cite{Sgaramella:2022}. The energy-calibration residual in the 4–12 keV interval is 1.5 eV. 
The long-term stability and rate dependence of the SDD energy scale are evaluated from repeated calibration runs performed throughout data taking, yielding contributions of 0.5 eV and 0.1 eV, respectively, to the systematic uncertainty.\\
A further contribution is associated with the treatment of higher kaonic deuterium transitions in the fit model. The fit includes the K-series components up to $6h \rightarrow 1s$; to assess the possible effect of unresolved additional transitions from higher-$n$ states on the extracted width and shift, alternative fits are performed by varying the number of kaonic deuterium transitions included in the model. The resulting systematic uncertainties are estimated to be 0.8 eV and 20 eV for the shift and width, respectively. \\
The sensitivity to the hadronic-background description is evaluated by replacing the nominal exponential function with a linear or a constant function, with freely varying parameters in each case. Systematic uncertainties of 1.1 eV on the shift and 26 eV on the width are assigned to this source.\\
The possible impact of the event selection on the shift and width is investigated by varying the cuts applied to the SDD time window, the Veto-1 timing selection and the charged-kaon-detector timing selection. The stability of the result under these variations is assessed using the Barlow test, by comparing the parameter shifts obtained from the nominal and modified selections with the corresponding statistical uncertainties.
No statistically significant dependence on the tested event-selection variations was observed. The individual systematic contributions are finally combined in quadrature to obtain the total systematic uncertainties on the measured kaonic deuterium shift and width.\\

\textbf{From shift and width to isospin-dependent $K^-N$ scattering lengths.}
Kaonic hydrogen and kaonic deuterium provide two independent experimental constraints on different combinations of the isoscalar $a_0$ and isovector $a_1$ components of the $K^-N$ scattering lengths.
The strong interaction induced shift $\varepsilon_{1s}$ and width $\Gamma_{1s}$ are related to the complex $K^-A$ scattering length, with $A=p,d$, through the summed-up Deser formula \cite{Baru:2009tx,Shevchenko:2021swf}:
\begin{equation}
\varepsilon_{1s}^{(A)}+\frac{i}{2}\Gamma_{1s}^{(A)}
=
\frac{
2\alpha^3\mu^2a_{K^-A}
}{
1+2\alpha\mu(\ln\alpha-1)a_{K^-A}
},
\label{eq:summed_deser}
\end{equation}
where $\alpha$ is the fine-structure constant, $\mu$ is the reduced mass of the system and the shift is defined as $\varepsilon_{1s} = E^{\mathrm{meas}} - E^{\mathrm{QED}}$, where $E^{\mathrm{meas}}$ is the measured transition energy and $E^{\mathrm{QED}}$ is the QED prediction. Comparisons with direct calculations indicate that the summed-up Deser formula has an accuracy better than approximately $2\%$ for kaonic hydrogen and $6\%$ for kaonic deuterium \cite{Shevchenko:2021swf}.\\
In the isospin-symmetric limit, the elementary charged-channel scattering lengths are related to $a_0$ and $a_1$ by
\begin{align}
a_{K^-p} &= \frac{1}{2}\left(a_0+a_1\right),\\
a_{K^-n} &= a_1.
\end{align}
Kaonic hydrogen therefore constrains one combination of $a_0$ and $a_1$, while kaonic deuterium provides a second relation involving a different combination:
\begin{align}
a_{K^-d}
=
\frac{4(m_N+m_K)}{2m_N+m_K}\,Q+C,\\
Q
=
\frac{1}{2}\left(a_{K^-p}+a_{K^-n}\right)
=
\frac{1}{4}\left(a_0+3a_1\right).
\label{eq:Kd_three_body}
\end{align}
where $m_N$ and $m_K$ are the nucleon and kaon masses. The term proportional to $Q$ represents the lowest-order impulse approximation, corresponding to kaon scattering from the individual nucleons. The term $C$ contains the higher-order contributions associated with the three-body $K^-pn$ dynamics, including charge exchange, multiple scattering, nucleon recoil and the deuteron structure. 
These contributions require a consistent few-body treatment, for example through Faddeev-type equations \cite{Shevchenko:2014uva,Shevchenko:2016wnu}.

\clearpage
\onecolumngrid

\begin{center}
    \textbf{\large Supplemental Material}
\end{center}
\label{sec:supplemental}

\begin{figure}[htbp]
    \centering
    \includegraphics[width=0.5\columnwidth]{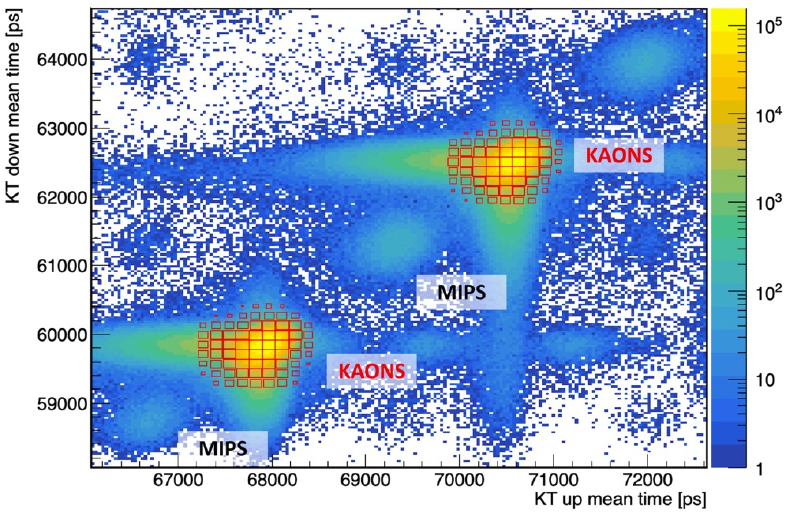}
    \caption{\textbf{Kaon trigger time-of-flight distribution.} Two-dimensional time-of-flight distribution with respect to the DA$\Phi$NE radio-frequency signal measured by the two scintillators of the kaon trigger (KT). The signals recorded by the upper and lower scintillators are plotted on the horizontal and vertical axes, respectively. The populations associated with charged kaons and minimum-ionizing particles (MIPS) are highlighted, showing a clear separation between true kaon events and MIPs.}
    \label{fig:kt_tof}
\end{figure}

\begin{figure}[htbp]
    \centering
    \includegraphics[width=0.5\columnwidth]{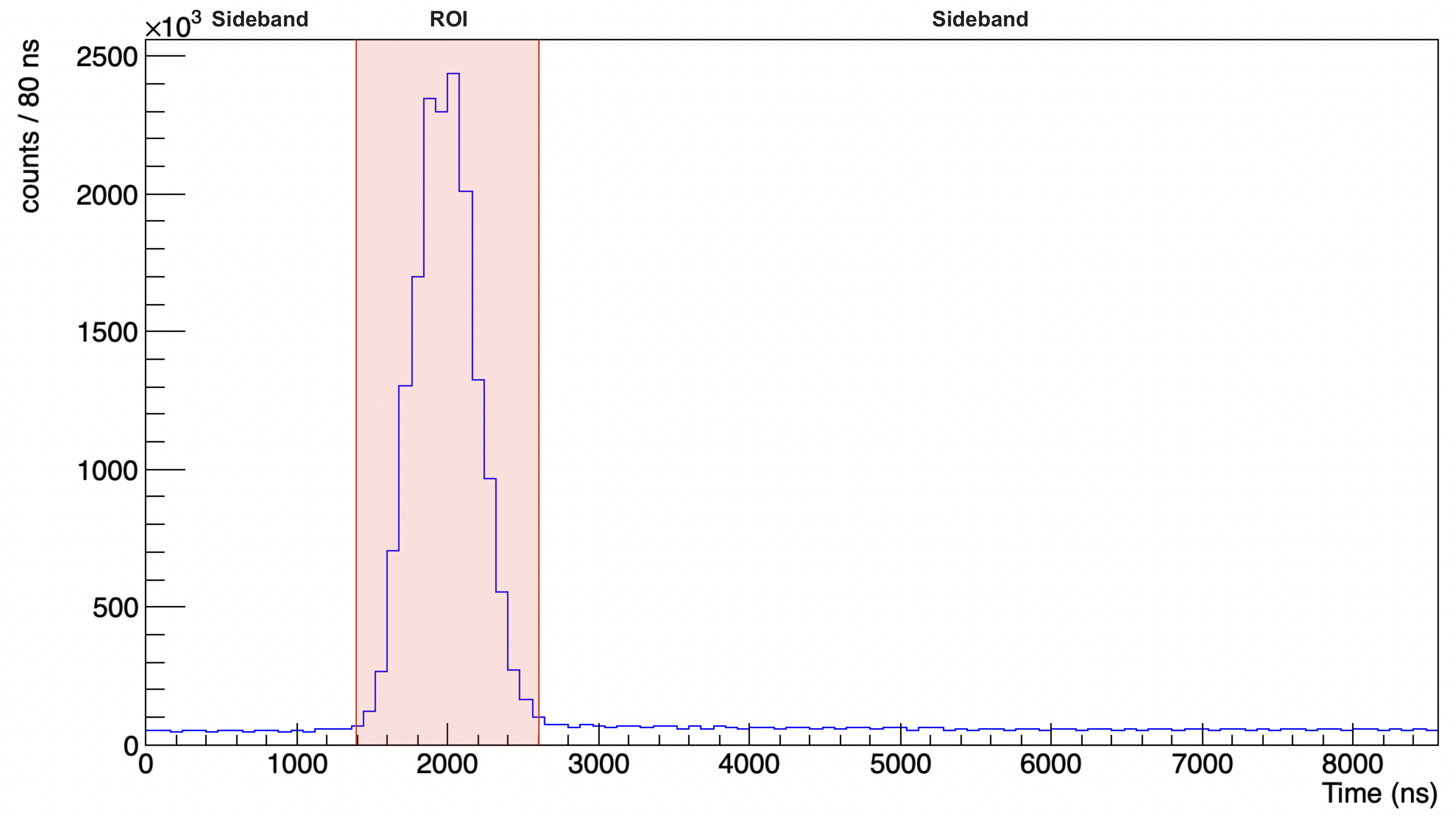}
    \caption{\textbf{SDD time distribution and event selection.} Time distribution of Silicon Drift Detectors (SDDs). The region of interest (ROI) used to select events synchronous with the kaon trigger is highlighted, together with the sidebands corresponding to electromagnetic background events.}
    \label{fig:dt}
\end{figure}

\begin{figure}[htbp]
    \centering
    \includegraphics[width=0.5\columnwidth]{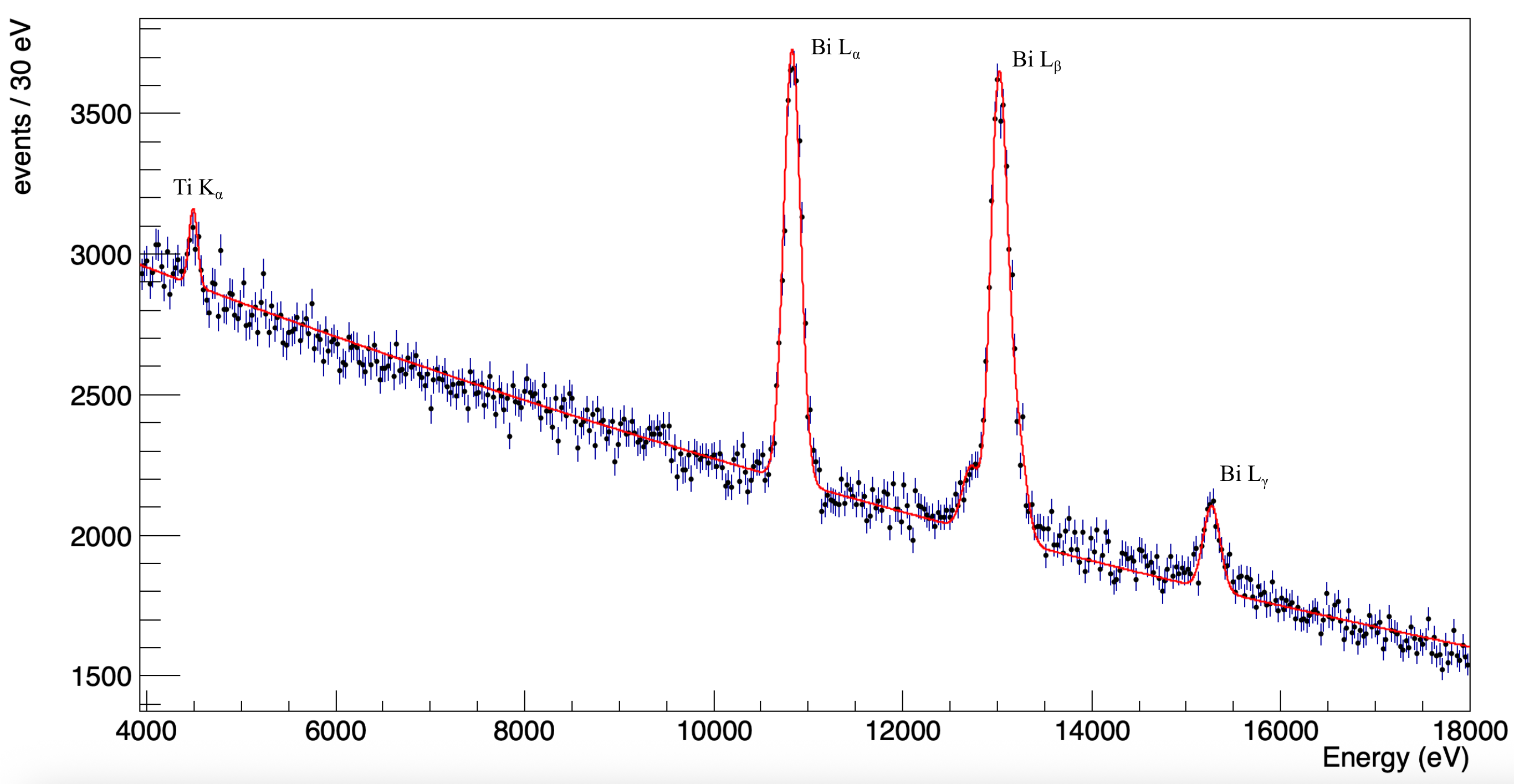}
    \caption{\textbf{Energy spectrum of the electromagnetic background.} Energy spectrum obtained after selecting events in the electromagnetic background region of the SDDs' time distribution. The data are fitted with an exponential function plus a constant term, together with Gaussian components describing fluorescence lines from the apparatus materials.}
    \label{fig:dt_energy}
\end{figure}

\end{document}